\documentclass[
  journal=pasa,
  manuscript=research-paper, 
  year=2025,
  volume=YY,
]{cup-journal}

\usepackage{amsmath, amssymb, microtype, siunitx, booktabs, gensymb, graphicx, subcaption, orcidlink, newunicodechar, sectsty, float, placeins}

\definecolor{pasablue}{HTML}{001074}

\usepackage{hyperref}
\hypersetup{
    pdftex,
    colorlinks=true,
    linkcolor=pasablue,
    citecolor=pasablue,
    urlcolor=pasablue,
    pdfborder={0 0 0}
}

\newcommand{\lt}{\ensuremath{<}}

\DeclareRobustCommand{\okina}{%
  \raisebox{\dimexpr\fontcharht\font`A-\height}{%
    \scalebox{0.8}{`}%
  }%
}
\newunicodechar{ʻ}{\okina}

\newcommand{\arcsec}{\mbox{$^{\prime\prime}$}}
\newcommand\farcs{\hbox{$.\!\!^{\prime\prime}$}}
\newcommand{\arcmin}{\mbox{$^{\prime}$}}
\newcommand{\Msun}{M$_{\odot}$}

\newcommand{\Lsun}{L$_{\odot}$}
\newcommand{\kms}{km~s$^{-1}$}
\newcommand{\Ha}{H$\alpha$}
\newcommand{\Hb}{H$\beta$}

\newcommand{\He}{He~}
\newcommand{\HeI}{He~{\sc i}}
\newcommand{\HeII}{He~{\sc ii}}
\newcommand{\NeII}{Ne~{\sc ii}}
\newcommand{\MgII}{Mg~{\sc ii}}
\newcommand{\CaII}{Ca~{\sc ii}}
\newcommand{\OI}{O~{\sc i}}

\newcommand{\OIII}{O~{\sc iii}}
\newcommand{\CIII}{C~{\sc iii}}
\newcommand{\CII}{C~{\sc ii}}
\newcommand{\CI}{C~{\sc i}}

\newcommand{\ion}[2]{#1\,\textsc{#2}}
\newcommand{\citefariasinprep}{Farias et al.\ (in prep.)}
\newcommand{\citefariasinprepalp}{Farias et al.\ in prep.}

\usepackage{amsmath}
\usepackage{amssymb,microtype,siunitx,booktabs,hyperref}
\usepackage[table]{xcolor}
\usepackage{tabularx}
\usepackage{booktabs}
\title{The Enigmatic Type Icn Supernova 2024abvb Located $\sim$22~kpc from Its Host Galaxy}

\author{J.~Shi}
\affiliation{OzGrav, School of Physics, The University of Melbourne, Parkville, VIC, Australia}
\email[J.~Shi]{jennifer.shi@student.unimelb.edu.au}

\author{K.~Auchettl}
\affiliation{OzGrav, School of Physics, The University of Melbourne, Parkville, VIC, Australia}
\alsoaffiliation{Department of Astronomy and Astrophysics, University of California, Santa Cruz, CA, USA}

\author{W.~B.~Hoogendam}
\affiliation{Institute for Astronomy, University of Hawai'i at Manoa, 2680 Woodlawn Dr., Hawai'i, HI 96822, USA}
\alsoaffiliation{NSF Graduate Research Fellow}

\author{D.~Farias}
\affiliation{DARK, Niels Bohr Institute, University of Copenhagen, Jagtvej 128, 2200 Copenhagen, Denmark}

\author{N.~Sarin}
\affiliation{Kavli Institute of Cosmology Cambridge, Madingley Road, Cambridge CB3 0HA, United Kingdom}

\author{K.~W.~Davis}
\affiliation{Department of Astronomy and Astrophysics, University of California, Santa Cruz, CA, USA}

\author{N.~Morrell}
\affiliation{Carnegie Observatories, Las Campanas Observatory, Casilla 601, La Serena, Chile}

\author{J.~T.~Hinkle}
\affiliation{Department of Astronomy, University of Illinois Urbana-Champaign, 1002 West Green Street, Urbana, IL 61801, USA}
\alsoaffiliation{NSF-Simons AI Institute for the Sky (SkAI), 172 E. Chestnut St., Chicago, IL 60611, USA}
\alsoaffiliation{Institute for Astronomy, University of Hawai'i at Manoa, 2680 Woodlawn Dr., Hawai'i, HI 96822, USA}
\alsoaffiliation{NHFP Einstein Fellow}

\author{D.~O.~Jones}
\affiliation{Institute for Astronomy, University of Hawai'i, 640 N. A'ohoku Pl., Hilo, HI 96720, USA}

\author{C.~Lidman}
\affiliation{The Research School of Astronomy and Astrophysics, Mt Stromlo Observatory, The Australian National University, Canberra, ACT 2611, Australia}
\alsoaffiliation{OzGrav, Centre for Gravitational Astrophysics, College of Science, The Australian National University, ACT 2601, Australia}

\author{C.~Angus}
\affiliation{Astrophysics Research Centre, School of Mathematics and Physics, Queen's University Belfast, Belfast BT7 1NN, UK}

\author{C.~Ashall}
\affiliation{Institute for Astronomy, University of Hawai'i at Manoa, 2680 Woodlawn Dr., Hawai'i, HI 96822, USA}

\author{C.~R.~Burns}
\affiliation{The Observatories of the Carnegie Institution for Science, 813 Santa Barbara St., Pasadena, CA 91101, USA}

\author{D.~D.~Desai}
\affiliation{Institute for Astronomy, University of Hawai'i at Manoa, 2680 Woodlawn Dr., Hawai'i, HI 96822, USA}

\author{A.~Do}
\affiliation{Institute of Astronomy, Madingley Road, Cambridge CB3 0HA, United Kingdom}
\alsoaffiliation{Kavli Institute of Cosmology Cambridge, Madingley Road, Cambridge CB3 0HA, United Kingdom}

\author{L.~Galbany}
\affiliation{Institute of Space Sciences (ICE-CSIC), Campus UAB, Carrer de Can Magrans, s/n, E-08193 Barcelona, Spain}
\alsoaffiliation{Institut d'Estudis Espacials de Catalunya (IEEC), 08860 Castelldefels (Barcelona), Spain}

\author{E.~Y.~Hsiao}
\affiliation{Department of Physics, Florida State University, 77 Chieftan Way, Tallahassee, FL 32306, USA}

\author{M.~E.~Huber}
\affiliation{Institute for Astronomy, University of Hawai'i at Manoa, 2680 Woodlawn Dr., Hawai'i, HI 96822, USA}

\author{M.~Y.~Kong}
\affiliation{Institute for Astronomy, University of Hawai'i at Manoa, 2680 Woodlawn Dr., Hawai'i, HI 96822, USA}

\author{B.~Martin}
\affiliation{The Research School of Astronomy and Astrophysics, Mt Stromlo Observatory, The Australian National University, Canberra, ACT 2611, Australia}

\author{K.~Medler}
\affiliation{Institute for Astronomy, University of Hawai'i at Manoa, 2680 Woodlawn Dr., Hawai'i, HI 96822, USA}

\author{A.~Möller}
\affiliation{Centre for Astrophysics and Supercomputing, Swinburne University of Technology, John St, Hawthorn, VIC 3122, Australia}
\alsoaffiliation{ARC Centre of Excellence for Gravitational Wave Discovery (OzGrav), John St, Hawthorn, VIC 3122, Australia}

\author{C.~Pfeffer}
\affiliation{Institute for Astronomy, University of Hawai'i at Manoa, 2680 Woodlawn Dr., Hawai'i, HI 96822, USA}

\author{A.~Polin}
\affiliation{Department of Physics and Astronomy, Purdue University, West Lafayette, IN 47907, USA}

\author{L.~Rauf}
\affiliation{The Research School of Astronomy and Astrophysics, Mt Stromlo Observatory, The Australian National University, Canberra, ACT 2611, Australia}
\alsoaffiliation{OzGrav, Centre for Gravitational Astrophysics, College of Science, The Australian National University, ACT 2601, Australia}

\author{S.~Romagnoli}
\affiliation{OzGrav, School of Physics, The University of Melbourne, Parkville, VIC, Australia}

\author{B.~Schmidt}
\affiliation{The Research School of Astronomy and Astrophysics, Mt Stromlo Observatory, The Australian National University, Canberra, ACT 2611, Australia}

\author{B.~J.~Shappee}
\affiliation{Institute for Astronomy, University of Hawai'i at Manoa, 2680 Woodlawn Dr., Hawai'i, HI 96822, USA}

\author{M.~D.~Stritzinger}
\affiliation{Department of Physics and Astronomy, Aarhus University, Ny Munkegade 120, DK-8000 Aarhus C, Denmark}

\author{A.~Syncatto}
\affiliation{Institute for Astronomy, University of Hawai'i, 640 N. A'ohoku Pl., Hilo, HI 96720, USA}

\author{B.~E.~Tucker}
\affiliation{The Research School of Astronomy and Astrophysics, Mt Stromlo Observatory, The Australian National University, Canberra, ACT 2611, Australia}
\alsoaffiliation{Australian National Centre for the Public Awareness of Science, The Australian National University, ACT 2601, Australia}

\author{M.~A.~Tucker}
\affiliation{Center for Cosmology and Astroparticle Physics, The Ohio State University, Columbus, OH, USA}
\alsoaffiliation{Department of Astronomy, The Ohio State University, Columbus, OH, USA}

\received {dd Mmm YYYY}
\revised  {dd Mmm YYYY}
\accepted {dd Mmm YYYY}
\published{22 September 202X}

\keywords{XXXXXXXXXXXXX} 
\begin{document}

\begin{abstract}
We report multiwavelength observations of the highly offset ($\sim$ 22.4 kpc) SN~2024abvb, the sixth Type~Icn supernova to date. With a peak magnitude of $\text{M}_{r}=-19.55\pm0.11$ mag, it is among the most luminous in the existing sample and shows similar colours and decline rates to other SNe~Icn. The early optical spectra show a blue continuum with narrow \CII\, features ($v_{\text{FWHM}} \sim2000\,$\kms), consistent with a typical wind velocity of a Wolf-Rayet star. The absence of \CIII\, $\lambda5696$ emission at the time of explosion is consistent with a Type Ibn supernova; however, the lack of narrow \He lines both in the optical and near-infrared spectra points towards a SNe~Icn classification. Unlike the majority of SNe~Icn, we do not detect broad features in the late-time ($7-21$ days relative to $o$-band peak) spectral phase of SN~2024abvb. Semi-analytical modelling of the light curves shows that it can be reproduced by $\sim2.6\,$\Msun\, of SN ejecta interacting with $\sim0.3\,$\Msun\, of circumstellar material (CSM); both larger than other SNe~Icn but consistent with rapidly evolving SNe~Ibn. The metallicity at the SN location is significantly lower than the global metallicity of its host galaxy, suggesting that line-driven mass loss, required to strip the progenitor of its H and He envelopes, was likely inefficient. We estimated the star formation rate history at the location of SN~2024abvb and find that it lies at the bottom $\sim5$th percentile among SESNe hosts, which is inconsistent with a Wolf-Rayet progenitor. Based on its spectral features, local and host environment properties, and host-galaxy offset, we favour an $8-10~\mathrm{M_\odot}$ star that is stripped by a compact companion as the progenitor that had a sufficient runaway velocity to reach the offset of SN~2024abvb.

\end{abstract}

\maketitle

\section{Introduction}
Stripped-envelope supernovae (SESNe) result from the core-collapse of massive stars $\gtrsim$~8\Msun, that have lost all or most of their outermost H (or He) layers \citep[e.g.,][]{Filippenko1997, Pastorello2007, GalYam2017, Modjaz2019}. Although the nature and evolution of SESNe progenitors remain uncertain\footnote{Note the recent discovery of hot ($T_{\rm eff}\sim50-100$\,kK), intermediate mass ($\sim2-8$\,M$_{\odot}$) helium stars whose hydrogen envelope has been stripped in a binary \citep{2023Sci...382.1287D, Gotberg2017}. Their properties are consistent with stars that were stripped by binary interaction and have an initial mass between $\sim8-25$\,M$_{\odot}$,  making them potential progenitors of SESNe.}, several mechanisms have been proposed to explain this envelope stripping. For example, radiative mass loss  by line-driven winds \citep[e.g.,][]{Heger2003, Smith2014}, strong stellar winds from a single, H-deficient massive star (i.e., a Wolf–Rayet (WR) star with M $\gtrsim 25$\Msun; e.g., \citealp{Yoon2017}), impulsive or eruptive mass loss \citep[e.g.,][]{Dessart2010, Moriya2014, Smith2014} or from mass transfer to a binary companion (e.g., \citealp{Podsiadlowski1992, Eldridge2013, Yoon2010, Gotberg2017, Ercolino2025}). In any of these scenarios, if the SN ejecta interacts with a dense circumstellar medium (CSM)\footnote{There is growing evidence that nearly 10\% of all core-collapse SNe interact with dense CSM \citep[e.g.,][]{Perley2020}.}, then the SN spectra during the photospheric phase will contain narrow emission lines -- H for SN IIn (e.g., \citealp{Schlegel1990}), He for SN Ibn (e.g., \citealp{Foley2007}, \citealp{Pastorello2008}), C/O for SN Icn (e.g., \citealp{Gal-Yam2022, Perley2022}), and Si/S for SN Ien (e.g., \citealp{Schulze2024}). Such SNe provide a unique opportunity to probe the diverse mass loss histories and CSM distributions associated with massive stars \citep[e.g.,][]{Fraser2020, Ercolino2025}.

Type Ibn SNe (SNe~Ibn) are a unique class of SESNe characterised by weak or absent H lines and narrow ($v_{\text{FWHM}}\sim2000$ \kms) He lines in their near-infrared (NIR) and optical spectra \citep[e.g.,][]{Foley2007, Pastorello2007, Shivvers2017, GalYam2017}. The narrow He lines are interpreted as evidence of shock interaction between the rapidly moving SN ejecta and He-rich, but H-depleted, CSM surrounding the progenitor \citep{Smith2017}. SNe~Ibn are rare, comprising only $\sim$1-2\% of core-collapse SNe \citep[e.g.,][]{Pastorello2008, Ma2025, Pessi2025}.
Since the discovery of SN~2006jc, the prototypical SN~Ibn, $\sim$80 SNe~Ibn have been spectroscopically classified on the Transient Name Server (TNS) \footnote{As of 2026 Feb 17}. 

Recently, population studies of SNe~Ibn have begun to probe their formation channels and attempt to explain the origin of their dense CSMs (e.g., \citealp{Hosseinzadeh2017, Maeda2022, Takatoshi2025, Farias2025}). Early studies of Type Ibn SN~2006jc suggested that massive WR stars ($\gtrsim25$\Msun) are possible progenitors of this class \citep{Foley2007, Pastorello2007, Tominaga2008}. However, studies of the environments of SNe~Ibn \citep[e.g.,][]{Hosseinzadeh2019, Dong2025, Warwick2025} and late-time observations of SN~2006jc identified a surviving companion (\citealp{Maund2016, Sun2020}), suggesting that (some) SNe~Ibn explode in binary systems. This is also further supported by theoretical studies that have attempted to explain the low-metallicity environments, and the spectral and CSM properties of SNe~Ibn using low mass He-stars found in binary systems \citep[e.g.,][]{Maund2016, Dessart2022, Wu2022, Ko2025} 

Type Icn SNe (SNe~Icn) constitute a recently identified subclass of SESNe, even rarer than SNe Ibn. Based on their discovery and analysis of SN~2019hgp, \citet{Gal-Yam2022} defined SNe~Icn as a distinct class, which had previously been theorised \citep{Smith2017, Woosley2017}. Only four additional SNe~Icn have been reported since SN~2019hgp (SNe~2019jc, 2021csp, 2021ckj, 2022ann; \citealp{Fraser2021, Perley2022, pellegrino_diverse_2022, Davis2023}). These SNe show narrow C, Ne, and O lines indicative of CSM interaction, while lacking H and He lines in their early spectra \citep{Gal-Yam2022}. More recently, \citet{Schulze2024} proposed a new SNe classification: Type~Ien, following the discovery of SN~2021yfj \citep{2024TNSAN.239....1G, 2024TNSAN.240....1G} whose spectra exhibit narrow Si, S, and Ar emission lines. \citet{2024TNSAN.239....1G} also proposed a potential intermediate class, Type~Idn, whose spectra would primarily show O, Ne, and Mg lines. However, no SN with Type~Idn spectral features have so far been discovered.

SNe~Ibn and Icn are characterised by fast light curve evolution, rising to their peak brightness in $\lesssim 10$~days \citep[][although some events have longer rise times, for example, the Type Ibn SN~2010al \citealp{Pastorello2015}]{Hosseinzadeh2017, Gal-Yam2022}. Their light curves typically reach peak luminosities of $r \approx -18.5$~mag. The rapid evolution suggests that the light curves are mainly powered by the SN shock interacting with a steep CSM profile \citep[e.g.,][]{Maeda2022, Nagao2023}. The photometric properties, including spectroscopic evidence for shock interaction between the ejecta and H-poor CSM, resemble the properties seen in the emerging class of extremely rare SN-like stellar explosions known as ``Luminous Fast Blue Optical Transients" (LFBOTs), which are thought to possibly result from the binary merger of a WR star with a companion (\citealp{Metzger2022}). However, if these SNe result from this evolutionary pathway, then it is thought that the delay time between the common-envelope phase and the merger is substantially longer than that inferred for LFBOTs \citep[see Fig. 4][]{Metzger2022}. Given the similarities between SNe~Ibn and SNe~Icn, it has been suggested that these events may represent a continuum of objects \citep[e.g.,][]{pellegrino_diverse_2022}, with their features gradually varying along a spectrum of progenitor composition and mass-loss history. As such, discovering and analysing SNe~Icn (in addition to SNe~Ibn) can uncover new insights into the final evolution of massive stars found in binary systems and their explosion properties, allowing us to place them in the context of the SESNe population. 

The origin of SNe~Icn remains uncertain. However, due to the similarity in features between SN~Ibn and SNe~Icn, it is plausible that these events arise from related progenitor systems, such as the core-collapse of a WR star. However, understanding the exact progenitor system is complicated by the wide range of spectral properties observed in these objects. Based on the properties of SN~2019hgp, \citealp{Gal-Yam2022} suggested that a spectroscopic subtype of WR stars that are C-rich and He-poor (WC stars) that underwent extreme mass loss before explosion are the progenitors of SNe~Icn (with SNe~Ibn arising from WR stars that are He and N rich - WN stars). From the inferred ejecta parameters, environments, and explosion properties, \citet{pellegrino_diverse_2022} suggested that there are multiple progenitor channels required to explain the diversity of properties associated with SNe~Icn. More specifically, they suggest that the progenitor of SN~2019jc is an ultra-stripped star (a massive star in a close binary system that has most or all of its He being stripped), while SNe~2019hgp, 2021ckj, and 2021csp were consistent with WR stars. These progenitor systems likely experienced an extreme mass-loss episode (e.g., \citealp{Perley2022} estimated a mass-loss rate of $0.5$\Msun~yr$^{-1}$ for SN~2021csp), stripping away their He and possibly C layers. Other studies have also suggested that WR stars undergoing extreme mass loss are not the only progenitor channel for these SNe. \citet{Nagao2023} suggested that SNe~2021ckj and 2021csp both potentially originate from a merger of a WR star with a compact object. This merger produces an explosion ejecta that includes an aspherical, high-energy component and a spherical ejecta component. Assuming this progenitor channel, the authors argue that despite their photometric similarities, the differences in their spectral features are attributed to the viewing angles of the jets based on the amount of CSM observed along the line of sight: SN~2021ckj was observed from the polar direction, whereas SN~2021csp was observed from an off-axis direction. Based on the low-metallicity environment, low ejecta masses and low CSM velocities, \citet{Davis2023} favour a binary-stripping progenitor scenario over a single WR progenitor for SN~2022ann. 

In the past two years, SNe with both Ibn and Icn spectral and photometric features have been discovered. For example, SN~2023emq \citep{Pursiainen2023} was initially classified as a SN~Icn due to the detection of a \CIII~$\lambda5696$ emission feature, also seen in SNe~Icn SNe~2019hgp and 2021csp. At eight days after peak brightness, its spectra developed strong, narrow \HeI\ emission typical of SNe~Ibn. However, some photometric properties were inconsistent with either class, leading \citet{Pursiainen2023} to tentatively conclude that SN~2023emq was a flash-ionised Type Ibn. Similarly, SN~2023xgo \citep{Gangopadhyay2025B,Yamanaka2025} showed an early narrow \CIII~$\lambda5696$ feature reminiscent of SNe~Icn, followed by \HeI\ emission after maximum light, resembling a SN~Ibn. \citet{Gangopadhyay2025} note that, if SN~2023xgo is interpreted as a flash-ionized SN~Ibn, its unusually strong carbon emission would render it unique within this class. More recently, aside form SN~2023emq and SN~2023xgo, \citet{Farias2025} identified eight additional Ibn-like SNe that also exhibit a clear \CIII~$\lambda5696$ feature (see Figure~A.3 in \citealt{Farias2025}). Due to their properties, these events offer novel insights into the extreme mass-loss histories and CSM distributions of massive star progenitors shortly before their death.

In this paper, we present an analysis of SN~2024abvb, one of the most luminous events among SNe~Ibn and Icn classes. Optical spectra obtained 7.9 days before peak show SNe~Ibn-like features (narrow \HeI\, and absent \CIII\, lines) while SNe~Icn features (\CII\, and \OI\, P Cygni profiles) emerge 7.2 days after peak. These observations indicate that SN~2024abvb exhibits characteristics of both SN~Icn and SN~Ibn. Previous observations of SN~2024abvb by \citet{Hu2026} suggest that its progenitor experienced extensive stripping of its outer layers prior to explosion. In addition,\citet{Anderson2026} report that SN~2024abvb exhibits a complex, nested structure in its CSM and is surrounded by a complex, nested CSM, indicating that the progenitor underwent multiple mass-loss episodes before core collapse. Here, we present a detailed multiwavelength analysis of SN~2024abvb and its host galaxy in the context of both the Type Icn and Ibn populations. This paper is structured as follows. In Section\,\ref{sec:obs} we present the photometric and spectroscopic observations and data reduction procedures. In Section \ref{sec:analysis}, we present our analysis of the host-galaxy environment, photometry, optical and NIR spectra, and compare these with SNe\,Ibn/Icn samples in the literature. We discuss our results and progenitor channels in Section\,\ref{sec:discussion} and conclude in Section\,\ref{sec:conclusion}.

Throughout the paper, we adopt a redshift $z = 0.0391 \pm0.0002$, as determined in \S\ref{sec:spec_properties}. This yields a distance modulus of $\mu= 36.2\,\pm\,0.01$~mag (where the uncertainty is determined from the redshfit uncertainty) or luminosity distance of $172\,\pm\,1$ Mpc (assuming a flat $\Lambda$CDM cosmology with $H_0=70$ km s$^{-1}$ Mpc$^{-1}$ and $\Omega_m=0.3$). We adopt a foreground Milky Way extinction of $E(B-V)_{MW} = 0.16$~mag \citep{Schlafly2011} toward SN~2024abvb, and all analyses are performed using Milky Way corrected light curves. We assume a negligible host extinction for SN~2024abvb owing to its large projected offset of $\sim$22.4\,kpc (see Section \ref{sec:host-galaxy} for more detail) from its host galaxy and due to the lack of a \ion{Na}{i}~D absorption line seen in our spectra (see Figure \ref{fig:hostgal}).

\section{Observations} \label{sec:obs}

\begin{figure*}[htbp]
  \centering
  \begin{minipage}{0.54\textwidth}
    \centering
    \includegraphics[width=\linewidth]{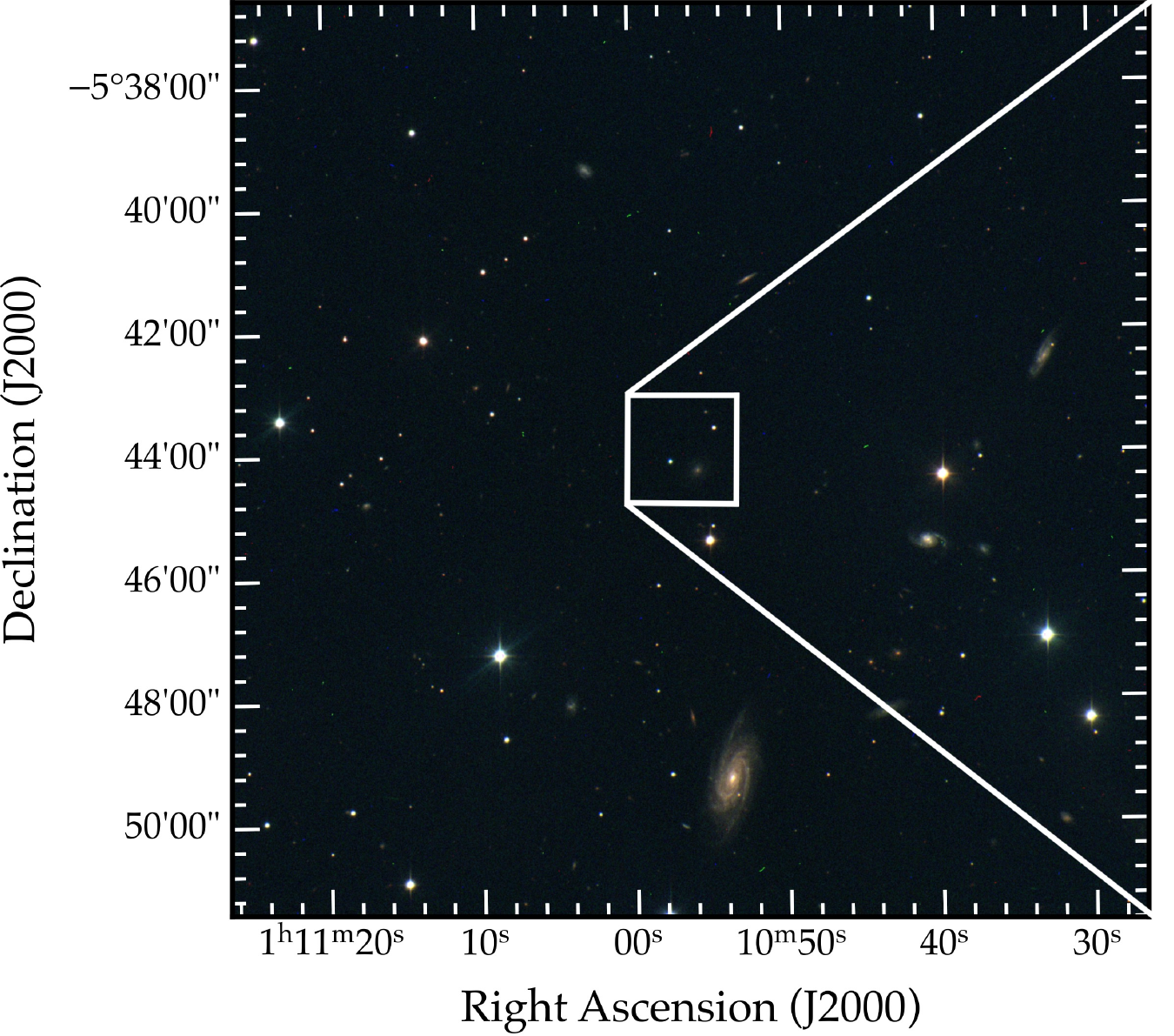}
  \end{minipage}
    \hspace{0.001\textwidth}
  \begin{minipage}{0.414\textwidth}
    \centering
    \raisebox{0.92cm}{%
        \includegraphics[width=\linewidth]{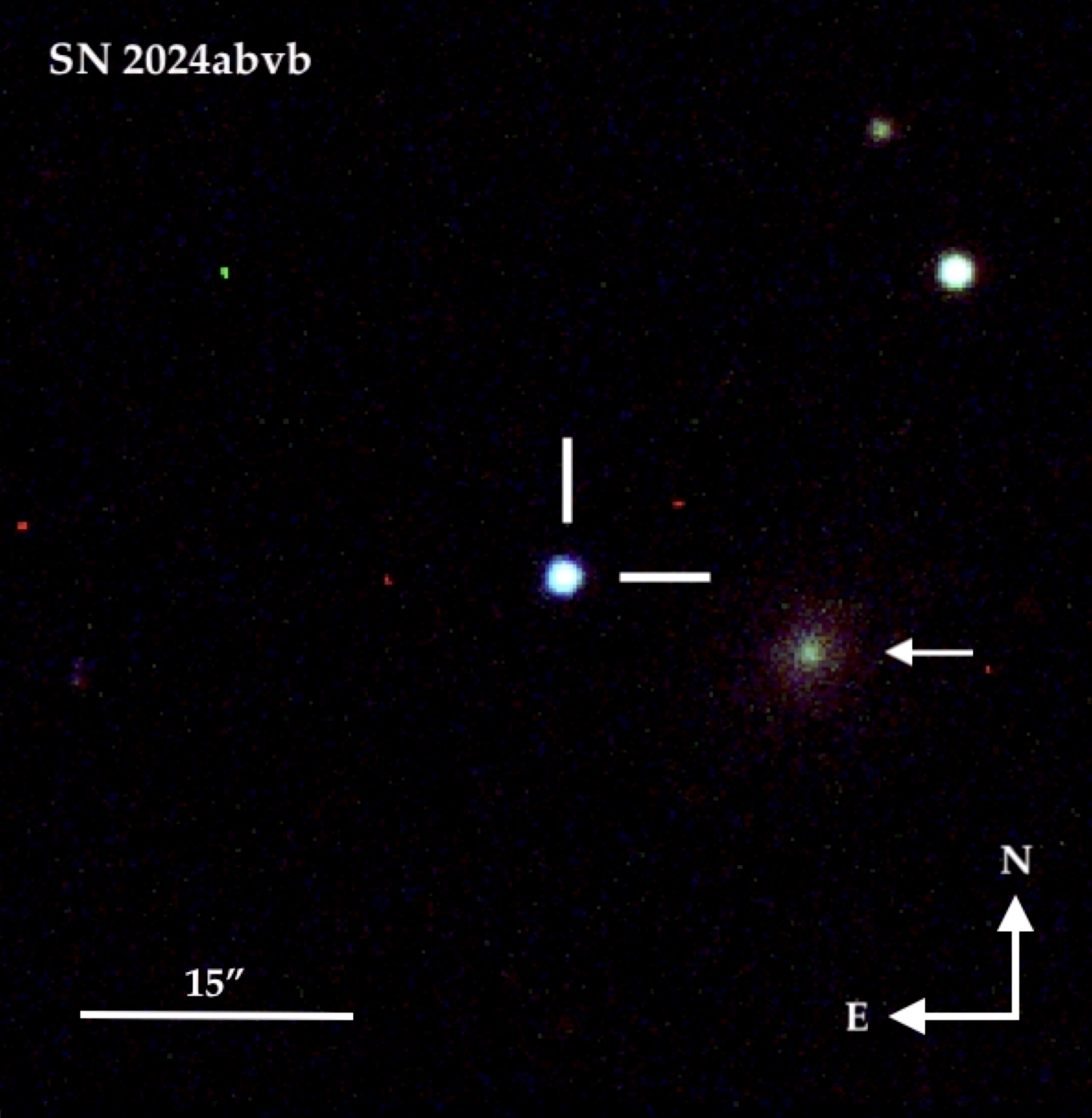}
    }
  \end{minipage}
  \caption{The left panel is a $15\arcmin\times 15\arcmin$ image of SN~2024abvb constructed using multi $Bgr$-band Swope images that are mapped to the blue, green, and red channels of the image, respectively. The right panel is a $2\arcmin\times 2\arcmin$ cutout with the position of SN~2024abvb marked. The arrow indicates the host galaxy PSO J011055.760-054416.73 with an apparent magnitude of $m_{AB}=18.27$ mag \citep{Flewelling2020}.}
  \label{fig:colorimage}
\end{figure*}

SN~2024abvb was discovered by the Asteroid Terrestrial-impact Last Alert System Project (ATLAS; \citealp{Tonry2018, Smith2020}) in the cyan ($c$) band with an apparent magnitude of $(18.78\pm0.05)$~mag on November 22, 2024 (MJD 60636.33) as ATLAS24qlh (see Figure \ref{fig:colorimage}). The last ATLAS non-detection ($c>19.4$~mag) occurred one day earlier, on November 21, 2024 \citep{Tonry2024}. The SN is associated with the host galaxy, PSO J011055.760-054416.73, with an offset of 22.4 kpc from the core of the host (see Figure \ref{fig:colorimage}; and Section \ref{sec:host-galaxy} for more discussion about the host properties). SN~2024abvb was classified on November 27, 2024, at UT 12:33:29 (6 days after discovery) as a SN Icn \citep{Stritzinger2024} based on a spectrum taken with the 2.56-m Nordic Optical Telescope (NOT) equipped with the Alhambra Faint Object Spectrograph and Camera (ALFOSC) as part of the NOT Unbiased Transient Survey.  

\subsubsection{Survey Photometry}
Photometry of SN~2024abvb was obtained from public ATLAS \citep{Tonry2018,Smith2020} and Zwicky Transient Facility (ZTF; \citealp{Bellm2019}) surveys. ATLAS data were retrieved from the ATLAS Transient Science Server \citep{Smith2020} with photometry derived using ATClean\footnote{\url{https://github.com/srest2021/atclean}} \citep{Rest2025}, which is based on procedures presented in \citep{Davis2023}. Likewise, ZTF observations were obtained through the ZTF forced photometry service \citep{Masci19, Masci23} and standard baseline corrections to the data and a signal-to-noise threshold of 3 were used, following the procedure outlined in the ZTF forced photometry manual v2.3\footnote{\url{https://irsa.ipac.caltech.edu/data/ZTF/docs/ztf_forced_photometry.pdf}}.
We also collected $riw$-band photometry of SN~2024abvb before and during the event from the Panoramic Survey Telescope and Rapid Response System (Pan-STARRS; \citealp[]{Kaiser2010, Chambers2016}) Survey for transients (PSST; \citealp[]{Huber2015}). Observations were performed using standard PSST procedures as outlined in \citet{Huber2015}. 

\subsubsection{Follow-up Photometry}\label{follow-up-phot}
Complementing the initial survey detections, we acquired targeted follow-up photometry to obtain additional multi-band measurements of SN~2024abvb.
The University of Hawai\okina i-owned and operated UKIRT 3.8m telescope observed SN~2024abvb with the NIR-imaging Wide Field Camera (WFCAM; \citealt{Casali07, Hodgkin09}). The data reductions follow those presented in \citet{Hoogendam2025b, Hoogendam2025a}. Briefly, the data were processed using \texttt{photpipe} \citep{Rest2005} and calibrated relative to 2MASS photometry of nearby field stars, following the methods described by \citet{Hodgkin09} and \citet{Peterson23}. We do not perform image subtraction when we run \texttt{DOPHOT} \citep{Schechter93} as part of \texttt{photpipe}. There is negligible host-galaxy flux at the location of SN~2024abvb because of its large offset. Therefore, we consider local background estimation and centroid determination from individual images sufficient to obtain reliable photometry.

Ultraviolet and optical photometry were obtained with the \emph{Neil Gehrels Swift Observatory} (\emph{Swift}; \citealp{Gehrels2004}) UV-Optical Telescope (UVOT; \citealp{Roming2005}) using the \citet{Poole2008} zero points updated by \citet{Breeveld10} and \citet{Breeveld11}. We used \texttt{HEASoft}\footnote{https://heasarc.gsfc.nasa.gov/docs/software/heasoft/} version 6.33.2 to obtain aperture photometry with the \texttt{UVOTSOURCE} package. An aperture of $5\arcsec$ is used for the source and a $20\arcsec$ source-free region located at ($\alpha_{J2000}, \delta_{J2000}$) = ($1^{\mathrm{h}}10^{\mathrm{m}}58^{\mathrm{s}}77, -05\degree{}43\arcmin26\arcsec58$) is used for the background. 

Additional optical photometric observations were taken as part of the Precision Observations of Infant Supernovae Explosions (POISE; \citealp{Burns2021_POISE}) Collaboration using the 1m~Swope telescope located at the Las Campanas Observatory (LCO). Photometric reductions follow the procedures described in \citet{Krisciunas2017}. 

\begin{figure*}[hbt!]
	\centering
		\includegraphics[width=\textwidth]{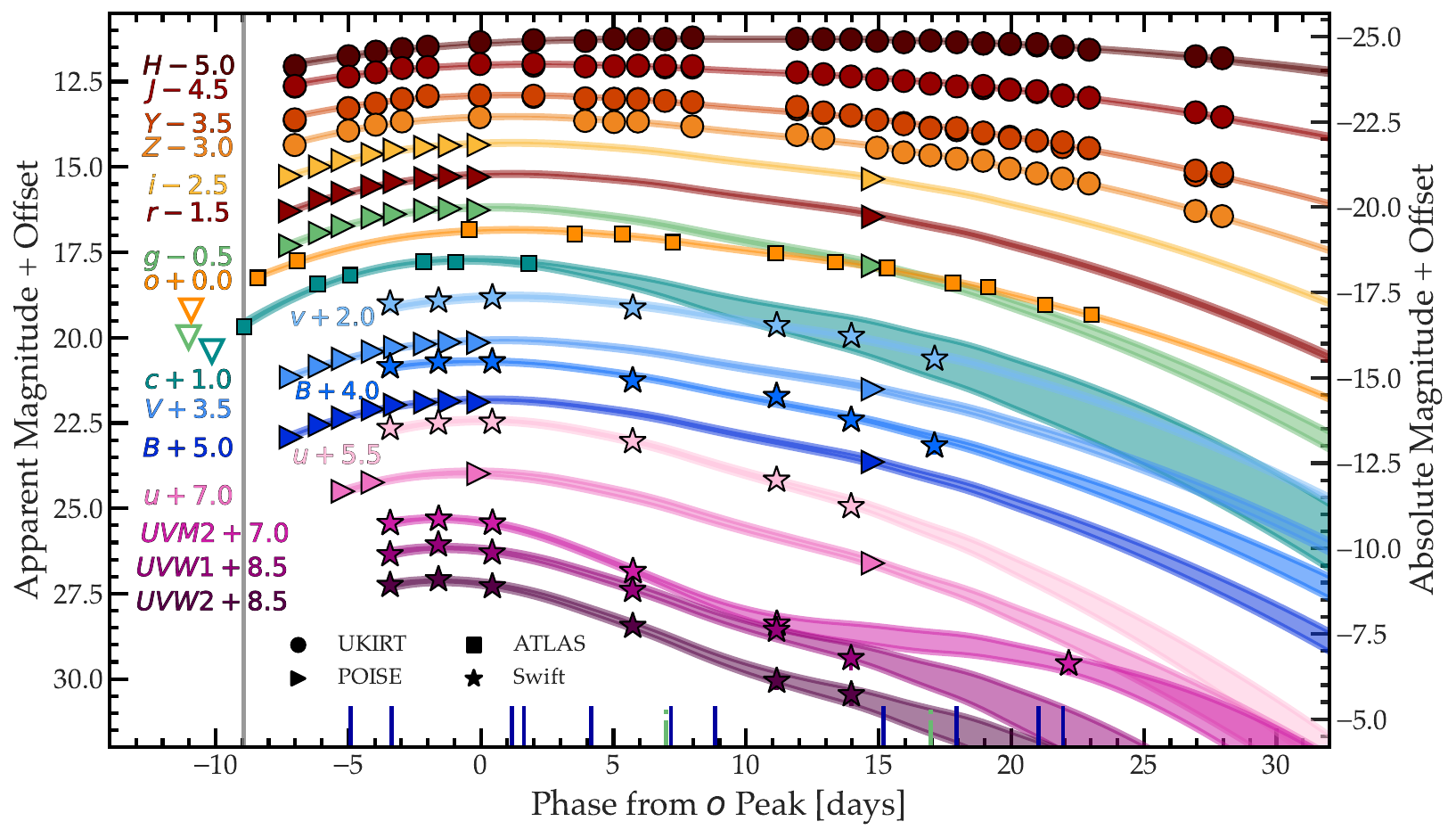}
	\vspace{-0.2cm}
    \caption{The UV, optical, and NIR light curves of SN~2024abvb. Upper limits are denoted with hollow triangles. The grey solid line indicates the time of discovery (MJD $60636.3$), which is $~3$ days after the estimated   time of explosion (MJD $60633.4$). The spectral epochs with optical (NIR) spectra are highlighted with solid blue (dashed green) lines. The phase is given relative to the $o$-band maximum (MJD $60645.3$). To guide the eye, shaded interpolation curves are overlaid for each filter band, indicating the Gaussian process fits within $1\sigma$. }
    \label{fig:lightcurves}
\end{figure*}

The complete multi-band light curve of SN~2024abvb is shown in Figure~\ref{fig:lightcurves}. While ATLAS was the first to discover SN~2024abvb, Pan-STARRS \citep{Chambers2016} detected the source with an observed apparent magnitude of $(19.9\pm0.05)$~mag (corresponding to an absolute magnitude of $(-16.2\pm0.05)$~mag) on MJD 60635.3 ($\approx1$ day prior to ATLAS's discovery) in the $w$-P1 filter. BlackGEM \citep{2024PASP..136k5003G} measured the SN about five days after (MJD 60640) in the BG-$q$-BlackGem filter at an apparent magnitude of $(17.5\pm0.02)$~mag (corresponding to an absolute magnitude of $(-18.9\pm0.02)$~mag). The last non-detections (denoted by empty coloured triangles in Figure \ref{fig:lightcurves}) prior to discovery are from ATLAS ($o > 19.2$ mag, 5$\sigma$; $c > 19.4 $ mag,  5$\sigma$) and ZTF ($g > 20.5$ mag, 5$\sigma$) which were obtained 10.9, 10.1, and 11.0 (MJD 60634.4, 60635.1, and 60634.3) days before $o$-band maximum brightness, respectively. 

\subsubsection{X-ray}\label{xray observations}
In addition to the \textit{Swift} UVOT observations, contemporaneous X-ray observations were obtained using \textit{Swift}'s X-ray Telescope \citep[XRT;][]{2005SSRv..120..165B} in photon-counting mode. All observations were reprocessed from level one XRT data using \texttt{XRTPIPELINE} version 0.13.7 with the most up-to-date calibrations, standard filters, and screening. Using a source region of radius $30\farcs0$ centred on the position of SN~2024abvb and a source-free background region of radius $150\arcsec.0$ located at ($\alpha_{J2000}, \delta_{J2000}$) = (1$^\mathrm{h}$\,11$^\mathrm{m}$\,10.6583$^\mathrm{s}$,\,$-5^\circ\,$39\arcmin\,31.574\arcsec), we find no significant X-ray emission associated with the SN in the individual epochs. To increase the signal-to-noise of observations, we merged all individual \textit{Swift} XRT observations using \texttt{XSELECT} version 2.5b and derived a 3$\sigma$ upper limit of 0.001 counts/second in the 0.3-10.0 keV energy range. Following \citet{Farias2025}, we assume an absorbed thermal Bremsstrahlung model with a temperature of 2.3 keV and a Galactic column density of $6.70\times10^{20}$ cm$^{-2}$ to convert this aperture-corrected count rate into a 3$\sigma$ upper limit on the flux (and luminosity). We obtain a 3$\sigma$ upper limit to the unabsorbed flux of $3.49\times10^{-14}$ erg\,cm$^{-2}$\,s$^{-1}$, which corresponds to a 3$\sigma$ upper limit to the unabsorbed X-ray luminosity of 1.26$\times10^{41}$ erg\,s$^{-1}$. 

\subsection{Spectroscopy}

Our optical spectra consist of two spectra from NOT, four spectra from the SuperNova Integral Field Spectrograph (SNIFS; \citealp{Lantz2004}) on the University of Hawai\okina i 2.2-meter (UH~2.2m) telescope as part of the Spectroscopic Classification of Astronomical Transients programme (SCAT; \citealp{Tucker2022c}), four spectra using the The Wide Field Spectrograph (WiFeS) that is mounted on the Australian National University 2.3 metre (ANU~2.3m) telescope (\citealp{Dopita07, Dopita2010, Price2024}), one spectrum from the Inamori Magellan Areal Camera and Spectrograph (IMACS) on the 6.5~m Magellan-Baade telescope \citep{Dressler2011}, and a final spectrum from the Keck-II~10.0~m telescope using the Keck Cosmic Web Imager \citep{Morrissey2018}.

SNIFS reductions used the pipeline presented in \citet{Tucker2022c}. The WiFeS spectra were taken in ``Nod \& Shuffle" mode, which produced simultaneous science and sky spectra, of which the sky contribution is subtracted during reduction \citep[see Section 2.2 of][for more details]{Carr2024}. The observations were reduced using the default reduction pipeline pyWiFeS \citep{Childress2014_pywifes} to produce calibrated 3D data cubes, of which 2D spectrum were extracted. NOT data were reduced following standard techniques using pyraf scripts within the \texttt{ALFOSCGUI} environment developed by E. Cappellaro. The flux calibration for the NOT spectra was performed using nightly sensitivity functions derived from standard star observations. An IMACS spectrum was obtained on Dec. 10, 2024 by A. Polin. Reductions were carried out through standard IRAF\footnote{The Image Reduction and Analysis Facility (IRAF) is distributed by the  National Optical Astronomy Observatory, which is operated by the Association of Universities for Research in Astronomy, Inc., under cooperative agreement with the National Science Foundation.} routines. Flux and telluric corrections were performed with appropriate standards observed during the same night. KCWI reductions were performed using the Keck Cosmic Web Imager Data Reduction Pipeline (KCWI DRP)\footnote{Accessible via \href{https://kcwi-drp.readthedocs.io/en/latest/index.html}{\url{https://kcwi-drp.readthedocs.io/en/latest/index.html}}.}. The flux calibration was based on spectrophotometric standard stars observed on the same night as SN~2024abvb.

Our NIR spectrum was taken on the 3.0m~NASA Infrared Telescope Facility (IRTF) with SpeX \citep{Rayner2003} in PRISM mode as part of the Hawai\okina i Infrared Supernova Study (HISS; \citealp{Medler2025a}). Reductions were performed with the Spextool \citep{Cushing2004}, including telluric correction using an A0V star, as in \citet{Hoogendam2025b, Hoogendam2025a}. 

A low-resolution spectrum covering the wavelength range 0.8-2.5 $\mu$m was obtained on 2024 December 8.04 UTC (MJD 60652.04) at an airmass of 1.1, with the Folded Port Infrared Echellette (FIRE) spectrograph on the 6.5 m Magellan Baade Telescope \citep{Simcoe2008}. The observation consisted of standard ABBA loops on the science target, totalling 507 s of on-target exposure time, followed immediately by an observation of the telluric standard HD4329. Detailed observational setup and reduction procedures are described by \citet{Hsiao2019}. A table of our spectroscopic observations is presented in Table \ref{tab:spectra_table}.

\section{Analysis} \label{sec:analysis}

\subsection{Host Galaxy Environment}\label{sec:host-galaxy}

We estimate the galactic properties of PSO J011055.760-054416.73, the host galaxy of SN~2024abvb, using the \texttt{Blast} tool \citep{Jones2024_Blast}\footnote{\href{https://blast.ncsa.illinois.edu/}{https://blast.scimma.org/}.}. \texttt{Blast} uses host-galaxy photometry and PSF-matched elliptical apertures from Galaxy Evolution Explorer \citep{Martin2005}, Sloan Digital Sky Survey \citep{Fukugita1996, York2000}, 2MASS \citep{Skrutskie2006}, Wide-field Infrared Survey Explorer \citep{Wright2010}, Dark Energy Survey \citep{Dey2019}, and Pan-STARRS \citep{Chambers2016}. The global spectral energy distribution (SED) properties are measured using \texttt{Prospector} \citep{Leja2019, Johnson2021, Wang2023}. 

The global properties of the host galaxy of SN~2024abvb are displayed in Table \ref{tab:host_properties_2024abvb}. The reported values and uncertainties are the median and 16th and 84th percentiles of the posterior distributions. The derived stellar mass ($\sim10^{9.7}\,\text{M}_\odot$) is below the mass of typical large galaxies ($10^{10}$\Msun$<$M$<10^{11}$\Msun) and above dwarf galaxies (M$<10^{8}$\Msun), which suggests it is an intermediate-mass galaxy. The host galaxy also has moderate star-formation ($\sim0.1\,\text{M}_{\odot}$yr$^{-1}$) and sub-solar metallicity ($\sim0.04\,$\text{Z}$_{\odot}$), approximately an order of magnitude lower to that of the Large Magellanic Cloud. We caution, however, that this metallicity estimate is derived solely from photometric measurements and may therefore be subject to significant uncertainty.

\begin{table}
\centering
\caption{Global properties of the host galaxy of SN~2024abvb.}
\label{tab:host_properties_2024abvb}
\begin{tabular}{lcc}
\toprule
Parameter & Value & Units \\
\midrule
Stellar Mass $M_{*}$ 
& $9.66^{+0.12}_{-0.12}$ 
& $\log_{10}(M_{\odot})$ \\

Star-Formation Rate (SFR) 
& $-0.99^{+0.80}_{-1.22}$ 
& $\log_{10}(M_{\odot}\,\mathrm{yr}^{-1})$ \\

Specific Star-Formation Rate (sSFR) 
& $-10.62^{+0.79}_{-1.22}$ 
& $\log_{10}(\mathrm{yr}^{-1})$ \\

Stellar Metallicity $Z_{*}$ \footnote{We note that this metallicity estimate is derived from photometric measurements alone and may not be robustly constrained. }
& $-1.38^{+0.38}_{-0.34}$ 
& $\log_{10}(Z_{\odot})$ \\

Mean Stellar Age 
& $7.93^{+3.78}_{-1.04}$ 
& Gyr \\
\bottomrule
\end{tabular}
\end{table}

\begin{figure*}[hbt!]
    \centering
    \begin{minipage}{0.49\textwidth}
    \centering
    \includegraphics[width=\linewidth]{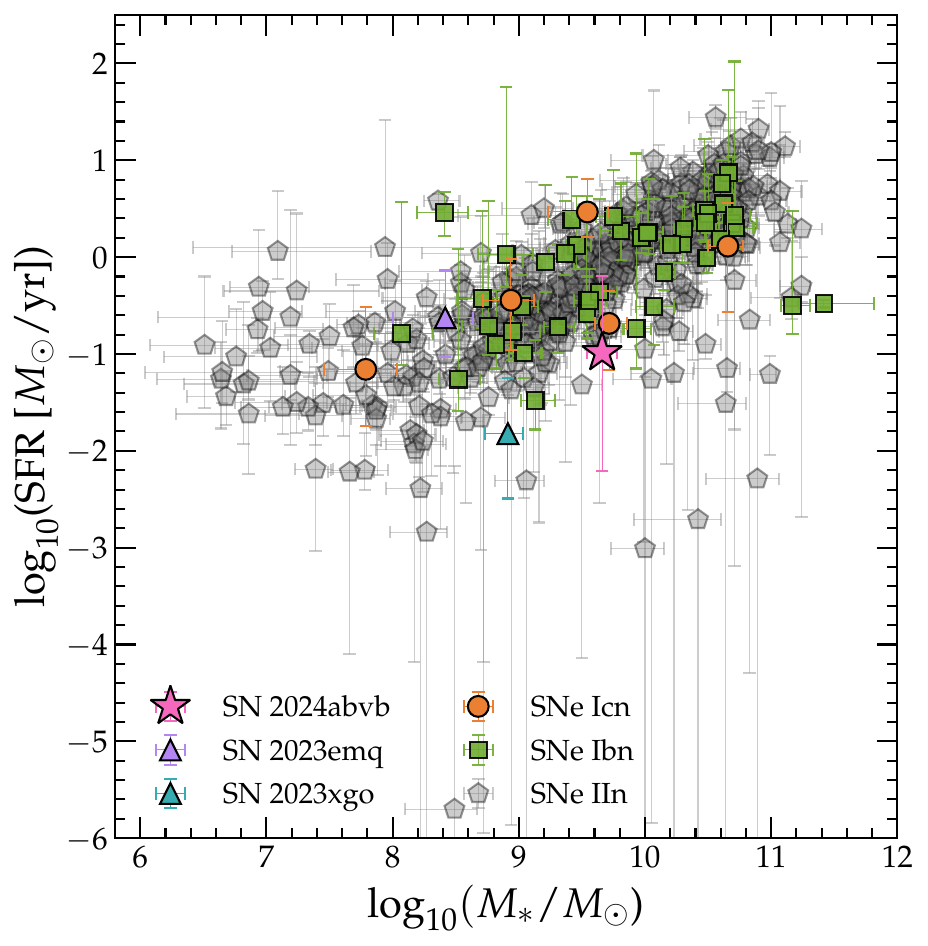}
    \end{minipage}
    \hspace{0.001\textwidth}
    \begin{minipage}{0.49\textwidth}
    \centering
    \includegraphics[width=\linewidth]{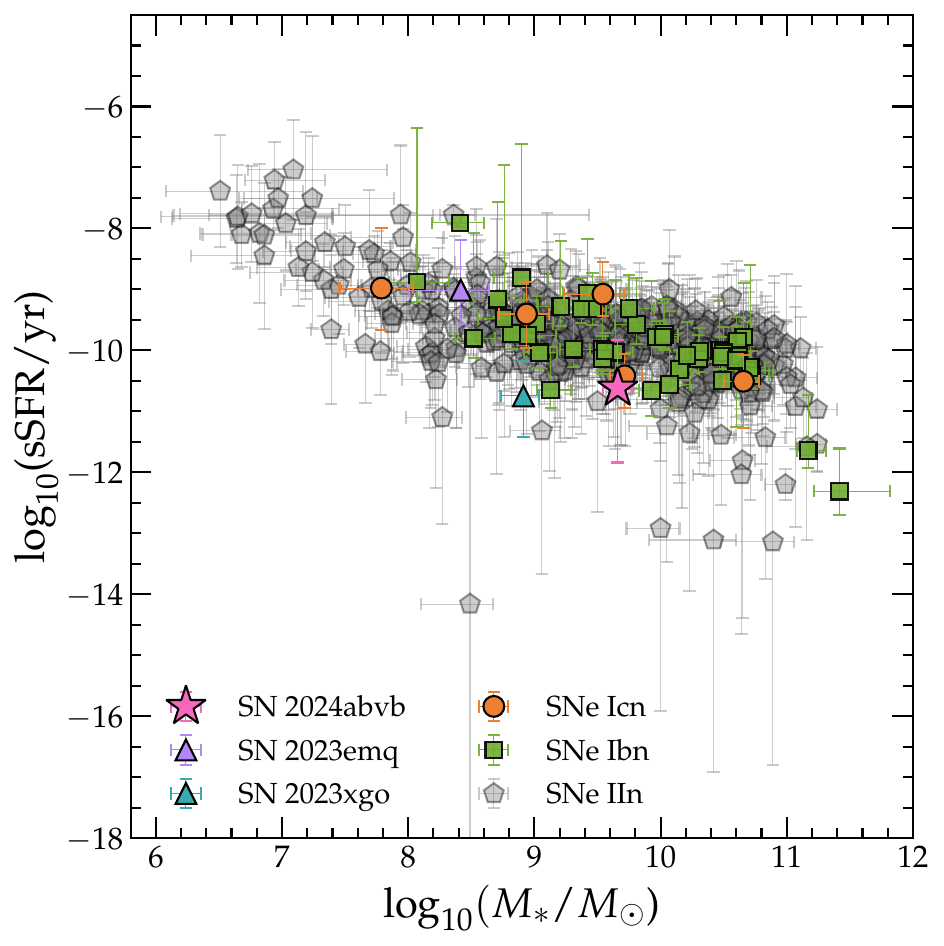}
    \end{minipage}
    \caption{Left: The host-galaxy stellar mass versus star formation rate (SFR) for SESNe and SNe~IIn. SN~2024abvb ($\log_{10}(\text{M}_{*}/\text{M}_{\odot}) = 9.66^{+0.12}_{-0.12}$) is denoted with the pink star. The SNe~IIn sample (taken from \citet{Schulze2021}) is shown as grey pentagons, the SNe~Ibn (taken from \citefariasinprepalp) are shown as green squares, and the SNe~Icn are shown as orange circles.  Using \texttt{Blast} \citep{Jones2024_Blast}, we also compare SN~2024abvb to SN~2023emq (purple triangle) and SN~2023xgo (blue triangle), the two transitional SNe~Ibn/Icn candidates found in the literature. Right: We compare the host-galaxy stellar mass and the Specific Star-Formation Rate (sSFR) for SESNe and SNe~IIn.}
    \label{fig:SFR_stellarmass}
    \label{fig:SFR_sSFR_stellarmass}
\end{figure*}

Figure \ref{fig:SFR_stellarmass} shows the SFR against stellar mass for thee hosts of SNe~Ibn (\citefariasinprepalp; \citealt{Sanders2013, Dong2025}; green squares), SNe~Icn (\citealp{Davis2023, GalYam2017, Perley2022, pellegrino_diverse_2022}; orange circles), SNe~IIn from PTF (\citealp{Schulze2021}; grey circles), and transitional SNe~Ibn/Icn candidates (\citealp{Pursiainen2023, Gangopadhyay2025B}; purple and blue triangles). Host-galaxy parameters for the SNe~IIn sample taken from \citet{Schulze2021} were extracted using \texttt{Prospector} (implemented in \texttt{Blast}), and the SNe~Ibn sample taken from \citefariasinprep\, also used \texttt{Blast}. Because most of the SNe~Icn and transitional Ibn/Icn candidates were originally modelled with different spectral energy distribution (SED) fitting methods, we re-derived their host-galaxy properties using \texttt{Blast} to ensure consistency across the full sample. The left panel of Figure \ref{fig:SFR_stellarmass} shows that the host galaxy of SN~2024abvb has a slightly lower SFR than the majority of SNe~Ibn and SNe~IIn hosts, while lying between the two transitional candidates. Its stellar mass is consistent with the median of galaxies hosting SESNe and SNe~IIn. In comparison to the SNe~Icn sample, SN~2024abvb has an SFR and stellar mass most consistent, within uncertainties, with SN~2021csp \citep{Perley2022}. The right panel of Figure \ref{fig:SFR_stellarmass} shows the relationship between the Specific Star-Formation Rate (sSFR) and stellar mass for our SNe samples. Overall, we find that the sSFR declines with stellar mass, indicating that more massive galaxies are less efficient at forming stars per unit stellar mass. At its given stellar mass, the host galaxy of SN~2024abvb lies toward the middle to lower end of the sSFR distribution, which may suggest that a progenitor outlived the primary episode of star formation or, statistically less likely, exploded before other high-mass stars formed \citep{Davis2023}. 

\begin{figure*}[hbt!]
    \centering
    \includegraphics[width=\linewidth]{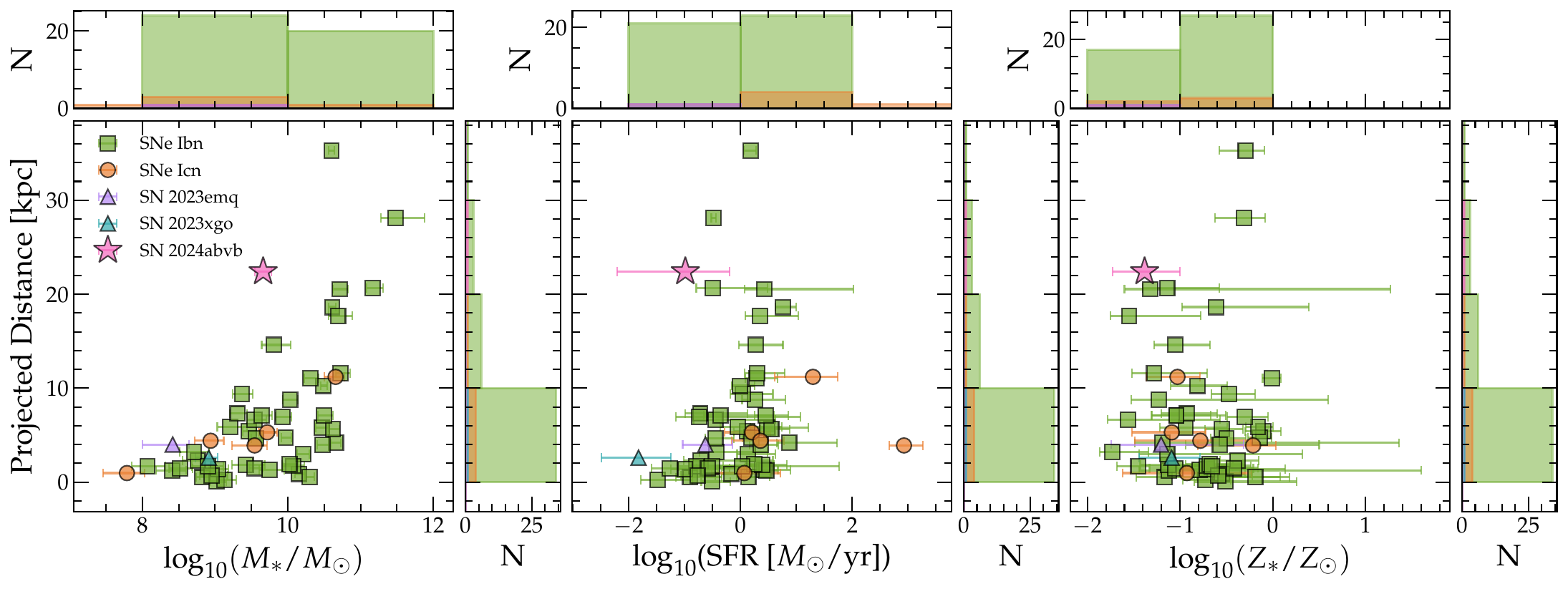}     
    \vspace{-2mm}

    \caption{Histograms and scatter plots of the host galaxy properties of interacting SESNe against projected distance. SNe~Ibn (green squares, 44) are collected from \citefariasinprep, \citet{Sanders2013} and \citet{Dong2025}, while the SNe~Icn (orange circles, 5), transitional SNe~Ibn/Icn candidates (purple and blue triangles), and SN~2024abvb (pink star) data are from \texttt{Blast} \citep{Jones2024_Blast}. The markers are the same across all three plots. The vertical uncertainty for SN~2024abvb is smaller than the plotted star, as the projected distance is derived from a distance modulus with a 0.1 mag uncertainty. The top histograms across the three plots show the number (N) of SNe per stellar mass, SFR, or stellar metallicity, respectively. \textit{Left}: host-galaxy stellar mass vs. projected distance. \textit{Middle}: SFR vs. projected distance. \textit{Right}: stellar metallicity vs. projected distance. SN~2024abvb has the largest projected offset amongst the SNe~Icn and transitional SNe~Ibn/Icn candidates, while its host-galaxy properties are broadly consistent with the SESNe population.}
    \label{fig:hist_comparisons}
\end{figure*}

Figure \ref{fig:hist_comparisons} shows the projected distance between the supernova and its host galaxy versus the stellar mass for our interacting SESNe sample. Interestingly, SN~2024abvb has the largest projected distance (22.4 kpc) compared to other SNe~Icn. SN~2024acyl \citep{Dong2025}, a SN~Ibn, has the (currently) largest known projected offset from its host galaxy, at $\sim$35kpc. A small number of other SNe~Ibn also have similar projected distances as SN~2024abvb, but they are hosted by larger host galaxies. Among the SNe~Ibn and SNe~Icn, there is a tentative trend that a larger stellar mass corresponds to a greater offset from its SN and this is supported by SN~2024abvb and SN~2024acyl. In the SNe-Icn sample, SN~2019jc has the second largest offset (11.2 kpc), which is half the projected distance of SN~2024abvb. The two SNe~Ibn with comparable offsets to SN~2024abvb are SN~2000er \citep{Pastorello2008} and LSQ13ccw \citep{Pastorello2015b}, but both events are associated with larger host galaxies ($\log_{10}(\text{M}_{*}/\text{M}_{\odot}) = 11.17^{+0.14}_{-0.09}$ and $\log_{10}(\text{M}_{*}/\text{M}_{\odot}) = 10.71^{+0.12}_{-0.09}$, respectively. 

At its projected distance, the host galaxy of SN~2024abvb exhibits a SFR lower than that of SNe~Ibn hosts, as shown in the middle panel of Figure \ref{fig:hist_comparisons}. In contrast, when compared with transitional candidates and SNe~Icn, SN~2024abvb shows the largest projected distance at its given SFR, however this is likely reflective of the small sample size of identified transitional sources and SNe~Icn. 

We compare the host-galaxy metallicity to the projected offset in Figure \ref{fig:hist_comparisons} (right). Again, we caveat that the metallicity estimates for the sample rely solely on photometric data and should be interpreted with caution due to significant uncertainties. The median metallicity values for all the interacting SESNe are slightly below solar metallicity, which suggests that a subsolar-metallicity host environment may be necessary to produce interacting SESNe, although we note that there are large uncertainties associated with our sample. SN~2024abvb has metallicity values consistent, within uncertainties, to $38\%$ of the SNe~Ibn population, and matches both the transitional SNe~Ibn/Icn candidates (i.e., SN~2023xgo ($\log_{10}(\text{Z}_{*}/\text{Z}_{\odot}) = -1.09^{+0.30}_{-0.34}$) and SN~2023emq ($\log_{10}(\text{Z}_{*}/\text{Z}_{\odot}) = -1.20^{+0.88}_{-0.53}$)), and two of the five SNe~Icn (i.e., SN~2019jc ($\log_{10}(\text{Z}_{*}/\text{Z}_{\odot}) = -1.03^{+0.24}_{-0.34}$), and SN~2021csp ($\log_{10}(\text{Z}_{*}/\text{Z}_{\odot}) = -1.09^{+0.37}_{-0.43}$)). If we instead use the galaxy mass–metallicity relation of \citet{Tremonti2004}, we find that the host galaxy of SN~2024abvb is estimated to have a metallicity of $12 + \log(\mathrm{O/H}) = 8.86$, corresponding to $\log_{10}(\text{Z}_{*}/\text{Z}_{\odot}) = 0.17$, assuming \citet{2009ARA&A..47..481A} abundances. Again, this is consistent within uncertainties with the host metallicities associated with our SESNe sample, but particularly the SESNe in our sample that are found in approximately solar-like metallicity environments. However, we note that given the large offset of SN~2024abvb from its host galaxy, the metallicity at the explosion site is likely lower than this global estimate.  
 
Overall, we find that despite its large projected offset to its host, the host of SN~2024abvb is consistent with the host properties of the population of interacting SESNe.

\subsection{Photometric properties and evolution}\label{photometricproperties}

SN~2024abvb has a similar rise and decline in all UV, Optical and NIR bands. We estimate the peak dates and magnitudes in the $o$- and $g$-bands using the Gaussian Process interpolation from the \texttt{PYTHON} package \texttt{george} \citep{Foreman-Mackey2015} with a Matern 3/2 kernel. The $o$-band maximum is estimated to be $60645.3\pm0.9$ MJD, while the $g$ band peaks at MJD $60644.0\pm1.5$. We define the rise time as the time between the explosion and the peak date of the $o$ band. Using the time of explosion as estimated in Section \ref{timeofexplosion}, the rise time of SN~2024abvb from explosion to maximum is $11.9^{+1.1}_{-1.2}$ days. This rise-time is comparable to that of SN~2022ann ($\sim$10 days; \citealt{Davis2023}), and consistent with two of the SNe~Ibn rise-times reported in \citet{Hosseinzadeh2017}, in particular PS1-12sk ($13.3\pm4.7$ days), SN~2014ak ($10.3\pm3.9$ days), and SN~2024acyl ($11.3\pm0.2$ days) \citep{Dong2025}. The rapid rise to peak brightness can be attributed to the CSM interaction, which can produce a fast increase in luminosity as the SN shock propagates through and expels the surrounding CSM and ejecta upwards \citep{Perley2022}.

Following Section 4.2 of \citet{Panjkov2024}, we search for evidence of pre-explosion variability from the progenitor of SN~2024abvb. We found no evidence for pre-explosion variability in the Pan-STARRS $w$, $r$, or $i$ bands for the progenitor of SN~2024abvb. The deepest constraint is provided by the $w$ band, which has the most extensive pre-explosion coverage, yielding a luminosity limit on the progenitor to be $\approx1.0\times10^{5}~$\Lsun\, (3$\sigma$).

\begin{figure*}[hbt!]
    \centering
    \includegraphics[width=\linewidth]{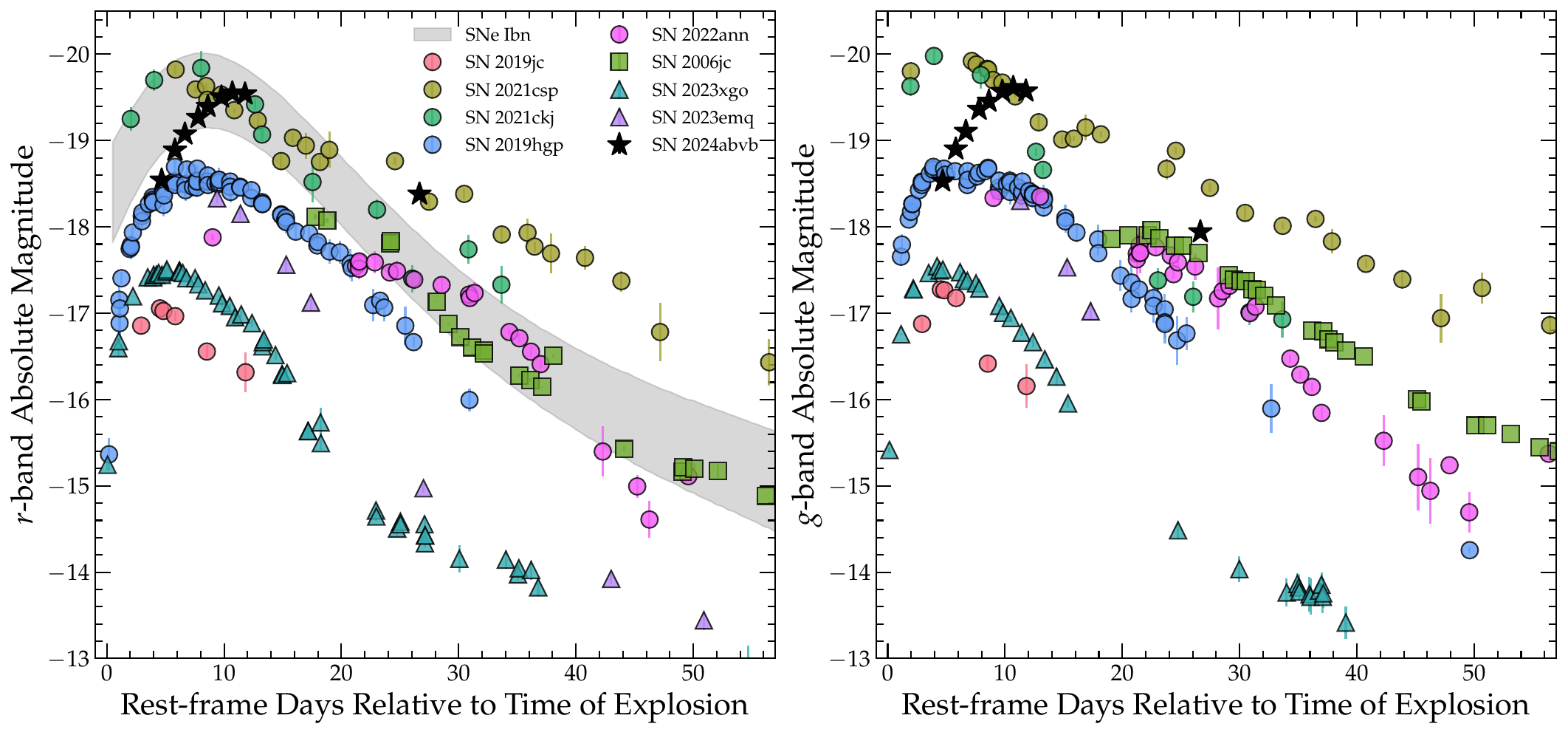}     
    \vspace{-2mm}
    \caption{$r$-band (left) and $g$-band (right) light-curve comparison of SN~2024abvb with a similar supernovae. Note that since the explosion time for SN~2006jc is not well-constrained, we align the estimated maximum-light epoch from \citet{Pastorello2007} to the maximum of the shaded SNe~Ibn template from \citet[shaded grey region]{Hosseinzadeh2017}. SN~2024abvb exhibits a rapid rise in its early evolution and is among the brightest sources. }\label{fig:grmag_comparison}
    \vspace{-2mm}
\end{figure*}

Figure \ref{fig:grmag_comparison} compares the $r$- and $g$-band absolute magnitude light curves of SN~2024abvb with other interaction-powered SESNe. Compared to the SNe~Icn sample, taken from \citep{Davis2023, pellegrino_diverse_2022, Gal-Yam2022}, SN~2024abvb is noticeably brighter in both bands ($M_r$ peak = $-19.55\pm0.11$ mag, $M_g$ peak = $-19.61\pm0.11$ mag), similar to SN~2021ckj and SN~2021csp. SNe~Icn show a wide range in peak magnitudes spanning$\sim3\,$mag, from the faintest event SN~2019jc to SN~2021csp, which reaches magnitudes comparable to those of superluminous SNe (e.g., \citealp[]{Angus2019}). SN~2024abvb is brighter than the canonical SN~Ibn, SN~2006jc, and has a similar shape in its rise to the SNe~Ibn template, although \citet{Hosseinzadeh2017} highlights that the template is biased towards brighter, slow-rising transients. In comparison to the two transitional SNe~Ibn/Icn sources (SN~2023xgo \citealp[]{Gangopadhyay2025B} and SN~2023emq \citealp[]{Pursiainen2023}) SN~2024abvb is roughly 1.2 (1.85) mag brighter in the $g$ ($r$) band.

SN~2024abvb declines at $0.07\pm0.007$~mag~day$^{-1}$ over two weeks in the $r$ band, assuming a constant post-peak decline despite the observational gap between $\sim 11$ and $+26$ days after explosion. Compared to other SNe~Icn, the decline rate (a week after peak brightness) of SN~2024abvb is similar to SN~2019hgp, SN~2019jc, and SN~2021ckj while roughly double the rate of SN~2022ann (10 days after its peak). This rate is also slightly lower than that of the SN~Ibn sample (0.1 mag day$^{-1}$ within 30 days post-maximum) \citep{Hosseinzadeh2017}. In contrast, SN~2023emq declines at roughly twice the rate ($\sim 0.2$ mag day$^{-1}$) over two weeks after peak, which \citet{Pursiainen2023} note is at the extreme end of SNe~Ibn/Icn. \citet{Hosseinzadeh2017} argue that such rapid decline rates ($0.05-0.15$ mag day$^{-1}$) rule out radioactive decay as a significant power source for the late-time light curves (except for a significant outlier, OGLE12-06) and instead supports circumstellar interaction as the most likely power source, consistent with the literature. 

SN~2024abvb exhibits a $g$-band evolution similar to that in the $r$ band, with a steep rise and a short-lived peak. Its $g$-band decline rate is 0.1 mag day$^{-1}\pm0.001$, similar to SN~2023xgo (0.14 mag day$^{-1}$), but nearly three times faster than the relatively flat decline of SN~2022ann (0.035 mag day$^{-1}$). 

\begin{figure}[hbt!]
    \centering
    \includegraphics[width=\linewidth]{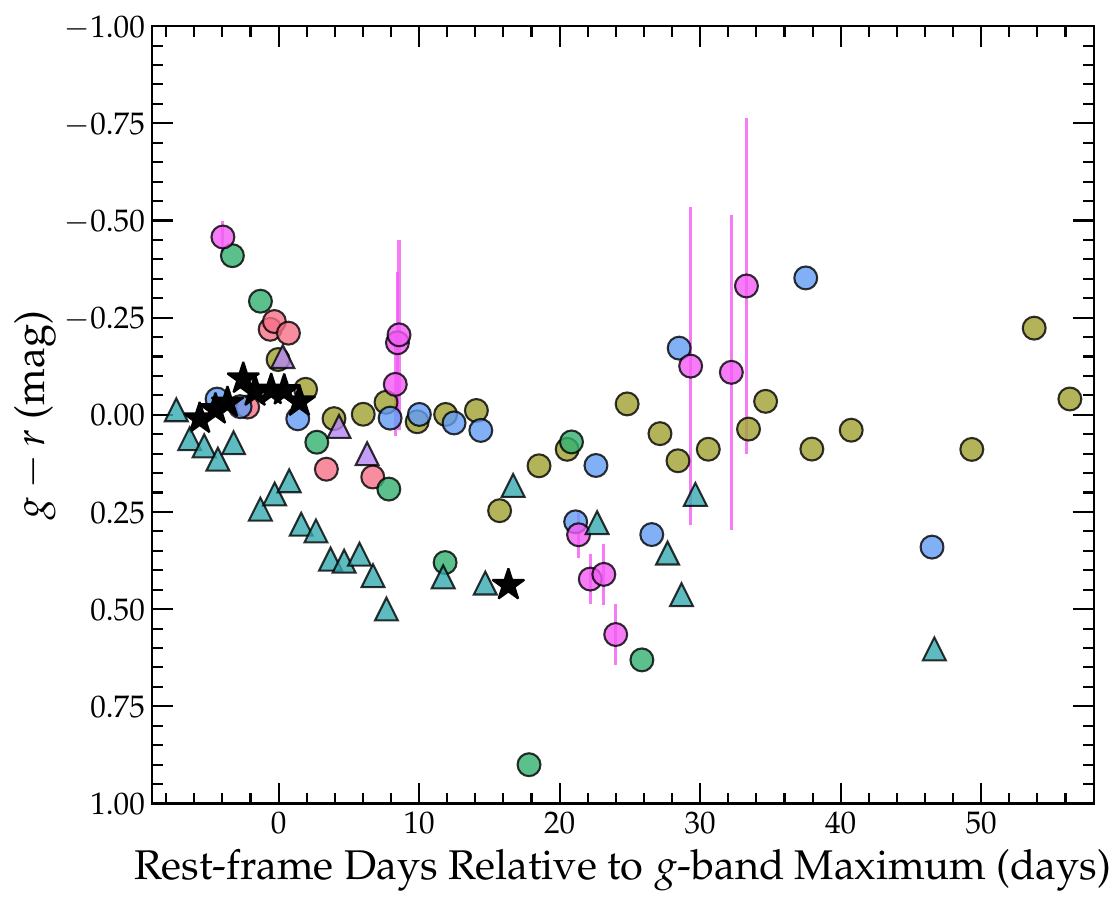}     
    \vspace{-2mm}
    \caption{The intrinsic colour evolution ($g-r$) of SN~2024abvb compared to the SNe~Icn \citep{pellegrino_diverse_2022, Davis2023}, transitional SNe~Ibn/Icn sample \citep{Gangopadhyay2025B, Pursiainen2023}. Markers are the same as Figure \ref{fig:grmag_comparison}. SN~2024abvb has a colour similar to that of both samples.}\label{fig:colour_comparison}
    \vspace{-2mm}
\end{figure}

In Figure \ref{fig:colour_comparison} we show the colour evolution of SN~2024abvb compared with the same SNe in Figure \ref{fig:grmag_comparison}. Near the maximum phase, SNe~Icn on average are bluer than the transitional sources. The colour evolution of SN~2024abvb remains consistently flat for the initial six days and grows redder over the subsequent fortnight. SN~2021csp and SN~2019hgp follow a similar evolution during their initial 30 days, while SN~2022ann begins significantly blue at maximum phase and grows redder. This broadly consistent colour evolution may indicate a long-lasting circumstellar interaction (CSI) powering source \citep{pellegrino_diverse_2022, Perley2022}. In contrast, SN~2023xgo and SN~2023emq show markedly redder colours during the first ten days (albeit with only three data points), differing from the evolution seen in SN~2024abvb. Overall, the colour evolution of SN~2024abvb exhibits features seen in both SNe samples. 

\subsubsection{Time of first light}\label{timeofexplosion}
Under the assumption that the first light occurs at the same time in all bands, we estimate the time of explosion for SN 2024abvb by fitting the ATLAS forced photometry to a two-component power-law model (similar to other studies, e.g., \citealp{Vallely2021, Fausnaugh2023, Hoogendam2024b, Hoogendam2025b, Hoogendam2025a}) defined as

\begin{equation}
    f(t) =
    \begin{cases}
      0 &  t < t_0 \\
       A\left( \frac{t - t_0}{1 + z} \right)^\alpha & t \geq t_0,
    \end{cases}
\end{equation}
\noindent where
\begin{equation}
    \alpha \equiv \alpha_1 \left (1 + \frac{\alpha_2(t-t_0)}{1+z}\right).
\end{equation}

\begin{figure}[hbt!]
    \centering
    \includegraphics[width=\textwidth]{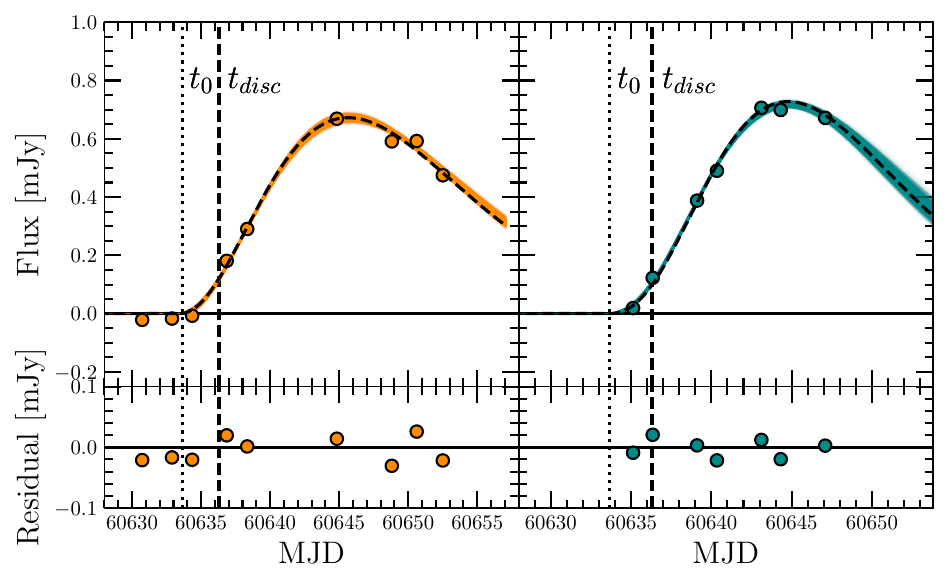}
    \vspace{-2mm}
    \caption{\textit{Left:} The ATLAS $o$-band rising light curve. \textit{Right:} The ATLAS $c$-band rising light curve. \textit{Bottom:} The fit residuals. The dotted line indicates the time of first light ($t_0$) from the best-fit parameters, and the vertical dashed black line indicates the time of discovery ($t_{\rm disc}$). The dashed black line denotes the best-fit light curve, and the coloured lines represent 1000 randomly sampled light curves from our posterior.}\label{fig:rising_lc_fits}
    \vspace{-2mm}
\end{figure}

The inferred value of $\alpha$ in the one-component model depends on the fraction of the rising light curve included in the fit \citep[e.g.,][]{Vallely2021}, but the $\alpha_1$ parameter from the two-component model remains insensitive. Rather, the parameters $\alpha_1$ and $\alpha_2$ correspond to the rise slope and the term necessary to trace the light curve peak curvature, which serves to minimise biases in $\alpha_1$ \citep[see][for more details]{Vallely2021}. $t_0$ is the time of first light, $A$ is a constant, and $z$ is the redshift. We fit the data using the \texttt{emcee} package (\citealp{ForemanMackey2013}), with results presented in Figure \ref{fig:rising_lc_fits} and Table \ref{tab:rising_lc_table}. 
The fitted time of explosion is MJD $60633.4^{+0.3}_{-0.2}$, $\sim$3 days before discovery (MJD$_{\rm discovery}=60636.33$), 4.0 (28.9) days before our first optical (NIR) spectrum, and 11.9 days before $o$-band maximum. 

\begin{table}[]
    \centering
    \caption{Best-fit parameters to the rising light curve of SN~2024abvb as shown in Figure \ref{fig:rising_lc_fits}.}\label{tab:rising_lc_table}
    \begin{tabular}{cccc}
        \hline
        Quantity & Value &  Units & Description \\
        \hline \hline
        \\    
            $t_0$ & $60633.38_{-0.22}^{+0.26}$     & MJD  & Time of first light    \vspace{0.50mm} \\
            $\alpha_{o{_1}}$ & $2.15_{-0.13}^{+0.14}$   &      & $o$-band rise slope    \vspace{0.50mm} \\
            $\alpha_{o{_2}}$ & $-0.024_{-0.001}^{+0.001}$  &      & $o$-band decline slope \vspace{0.50mm} \\
            $A_o$ & $0.01_{-0.01}^{+0.01}$         & mJy  & $o$-band constant      \vspace{0.50mm} \\
            $\alpha_{c{_1}}$ & $2.64_{-0.19}^{+0.21}$   &      & $c$-band rise slope    \vspace{0.50mm} \\
            $\alpha_{c{_2}}$ & $-0.02_{-0.01}^{+0.01}$  &      & $c$-band decline slope \vspace{0.50mm} \\
            $A_c$ & $0.01_{-0.01}^{+0.01}$         & mJy  & $c$-band constant      \vspace{0.50mm} \\
 \vspace{0.50mm} \\
            \hline \\
    \end{tabular}
\end{table}

\subsubsection{Bolometric Analysis}
From our multiwavelength light-curve coverage, we estimate the bolometric properties using \texttt{extrabol} \citep{Thornton2024}. After a series of blackbody SEDs is fit to the broadband light curves, \texttt{extrabol} infers the luminosity, photospheric radius, and temperature evolution.  

\begin{figure}[hbt!]
    \centering
    \includegraphics[width=\linewidth]{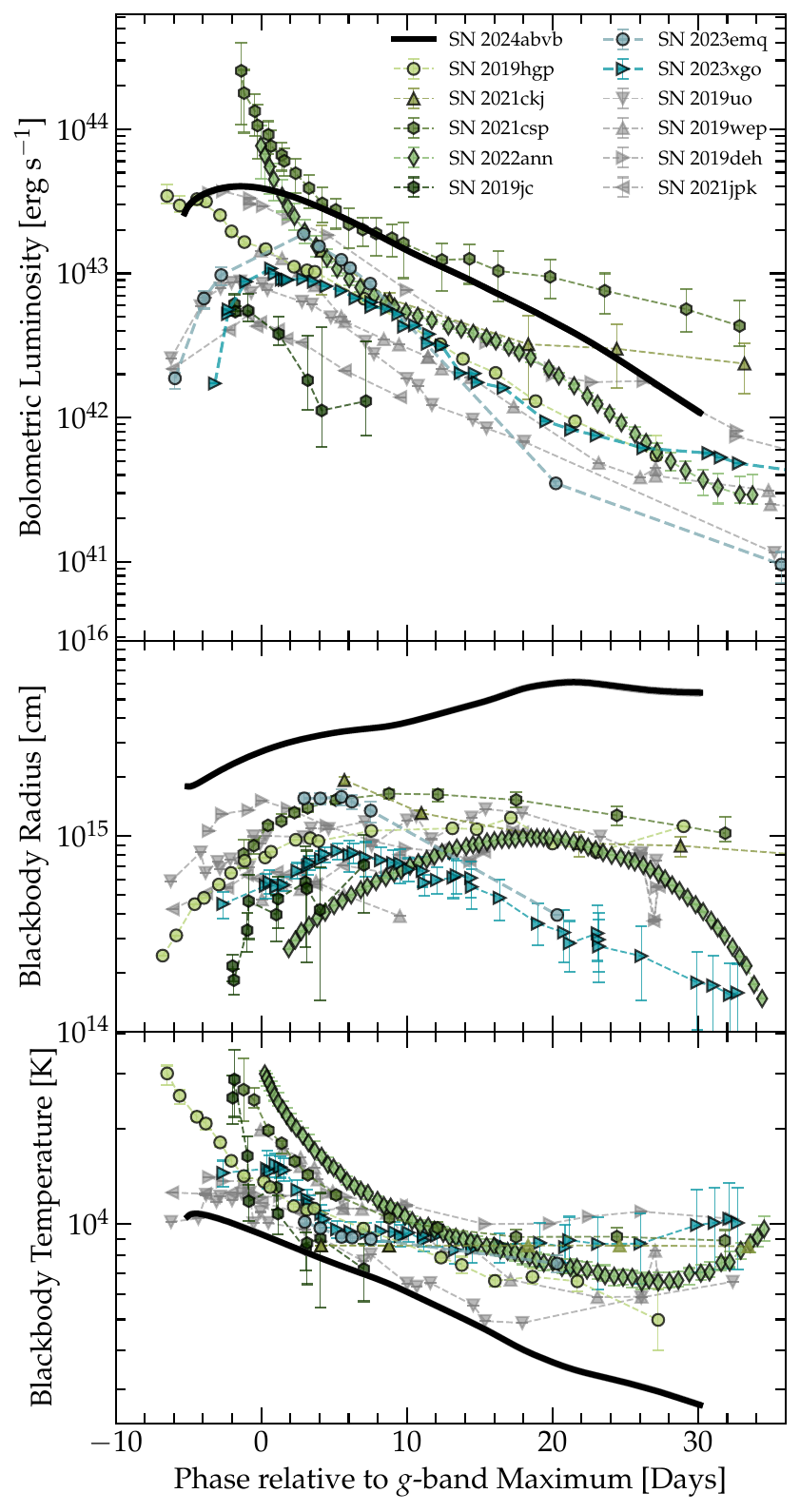}     
    \vspace{-2mm}
    \caption{The estimated bolometric light curve (top panel), blackbody radius evolution (middle panel), and blackbody temperature evolution (bottom panel) of SN 2024abvb (black curves with the 1$\sigma$ uncertainty plotted as a grey band but not visible), a sample of SNe~Icn (green markers;  \citealp[]{pellegrino_diverse_2022, Perley2022, Davis2023}), SNe~Ibn (grey markers; \citealp[]{Pellegrino2022}), and transitional SNe~Ibn/Icn candidates (blue markers; \citealp[]{Pursiainen2023, Gangopadhyay2025B}). Note that SN~2022ann, SN~2021ckj, and SN~2019jc did not have UV observations. The grey shaded band in the top panel is the template light curve of SNe~Ibn from \citet{Hosseinzadeh2017}. The phase is relative to the $g$-band maximum. The bolometric luminosity of SN~2024abvb appears consistent with that of the comparison SNe sample; however, it exhibits a significantly larger blackbody radius and a substantially cooler temperature.}\label{fig:bbluminosity}
    \vspace{-2mm}
\end{figure}

Figure \ref{fig:bbluminosity} presents the bolometric light curve and the blackbody radius and temperature evolution of SN~2024abvb, based on 15 photometric light curve epochs spanning the UVOT UV to the UKIRT IR bands. The bolometric luminosity of SN~2024abvb follows an approximately linear decline from $\sim$6-30 days after explosion. SN~2024abvb rises to a peak of $\sim$11,000 K within six days and rapidly cools over the next 25 days to approximately $4000$ K.
SN~2024abvb exhibits a blackbody radius evolution not seen in other SNe~Icn or Ibn. Its radius rapidly shrinks when it reaches its maximum temperature and then gradually expands to $6.1\times10^{15}$\,cm at 32 days after the time of explosion. From there, the radius plateaus to a radius of $\sim(5-6.1)\times10^{15}$\,cm. 

The top panel of Figure \ref{fig:bbluminosity} shows a comparison of SN~2024abvb's bolometric luminosity with our SNe~Ibn and SNe~Icn samples. About three days before maximum, SN~2024abvb has a peak luminosity of $4\times10^{43\,}\text{erg}~\text{s}^{-1}$, making it one of the brighter sources in our sample, similar to SN~2019hgp, SN~2022ann, and SN~2021csp ($\sim3\times10^{43}\,\text{erg}~\text{s}^{-1}$, $\sim4\times10^{43}\,\text{erg}~\text{s}^{-1}$, $\sim3.5\times10^{43}\,\text{erg}~\text{s}^{-1}$, respectively). The bolometric luminosity evolution of SN~2024abvb is similar to that of SNe~Ibn and transitional events. Its rise to peak brightness and decline until $\sim25$ days after peak closely follows the SNe~Ibn template (\citet{Hosseinzadeh2017}; grey band). Although the transitional SNe~Ibn/Icn candidates and selected SNe~Ibn sources are less luminous, they exhibit comparable luminosity evolution. After $\sim25$ days, SN~2024abvb fades quickly, resembling SN~2022ann \citep[Type Icn;][]{Davis2023}.  

The middle panel of Figure \ref{fig:bbluminosity} compares the blackbody radius of SN~2024abvb with the interacting SNe sample. SN~2024abvb has the largest photospheric radius of all SNe Ibn/Icn in our sample. Its evolution is akin to SN~2022ann from 5 days before peak, in which the radii begins to recede much later ($\sim 20$ days after peak brightness) compared to the other objects that recede within a week after the maximum. During the first three weeks after peak luminosity, the estimated blackbody radius, corresponding to the photosphere, expands at a constant velocity of $18,600$~\kms, after which it stays constant. However, this value is extremely high, given that the maximum velocity we detect from the spectral features is only $~2000$~\kms~ (see Figure \ref{fig:CII_timeseries}), indicating that the spectral features trace the CSM environment. 

The temperature evolution of SN~2024abvb is cooler than that of all comparison SNe samples (see the bottom panel of Figure \ref{fig:bbluminosity}). Unlike the majority of interacting SNe that are initially very hot (e.g., T$_{\text{BB}} > 25,000$ K for SN~2022ann), SN~2024abvb has an initial temperature of $\sim10,500$ K. After two days, SN~2024abvb reaches a maximum temperature of $\sim10,800$ K and cools over the next 45 days. This behaviour contrasts with that of the SNe~Ibn sample, which show a temperature increase over approximately 5 days before maximum light, followed by a faster subsequent decline. Early-time temperatures for SNe~Icn and the transitional candidates decrease from the earliest epochs and begin to plateau around a week after maximum light, cooling more slowly than SN~2024abvb. 

In general, the evolution of the blackbody properties is broadly similar to that of the other SNe~Ibn in our sample, which could suggest that the ejecta and circumstellar environment of SN~2024abvb is more similar to this population rather than our SNe~Icn sample.

\subsubsection{Photometric Modelling}

\begin{figure*}[htbp]
  \centering
  \begin{minipage}{0.45\textwidth}
    \centering
    \includegraphics[width=\linewidth]{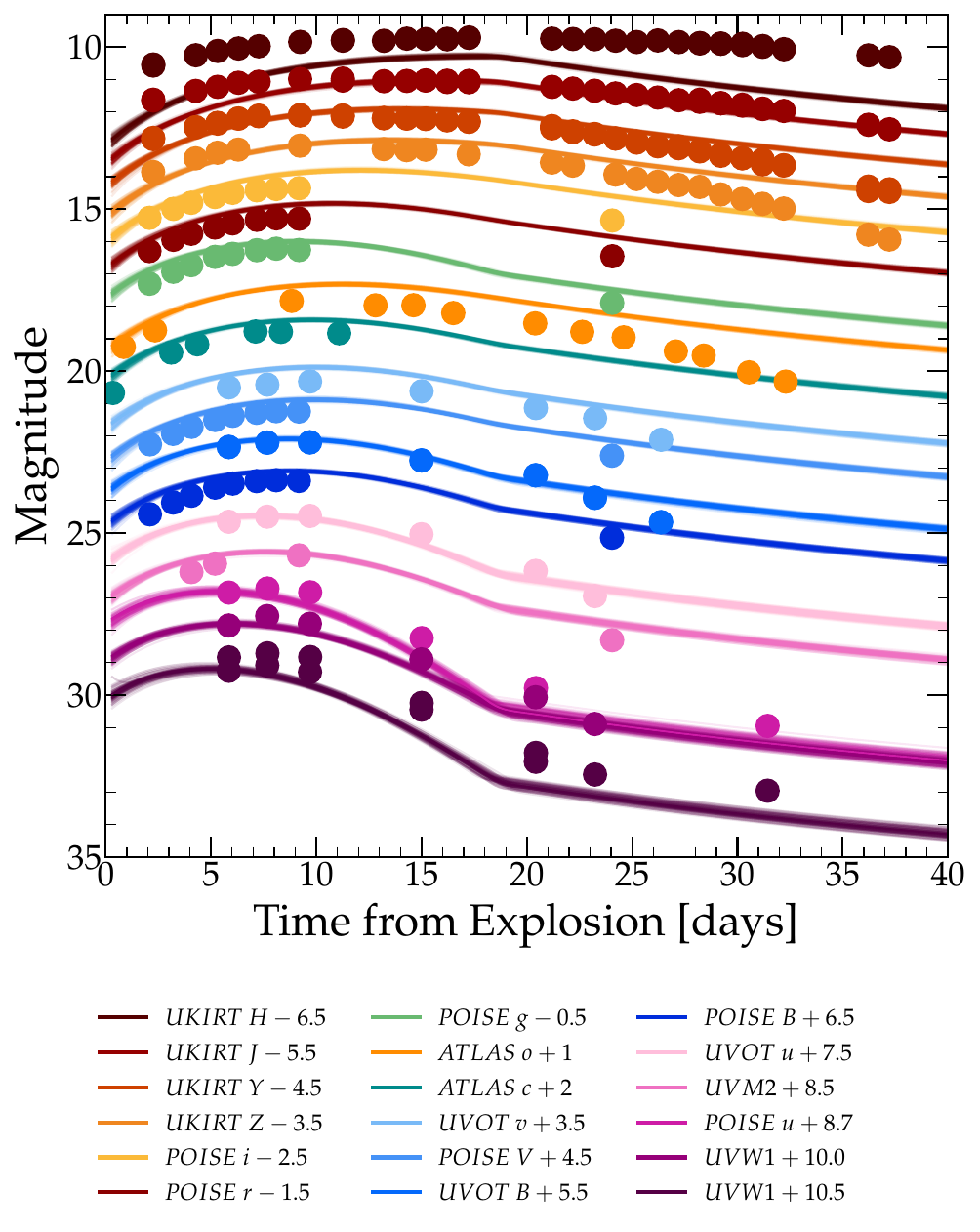}
  \end{minipage}
    \hspace{0.001\textwidth}
  \begin{minipage}{0.54\textwidth}
    \centering
    \includegraphics[width=\linewidth]{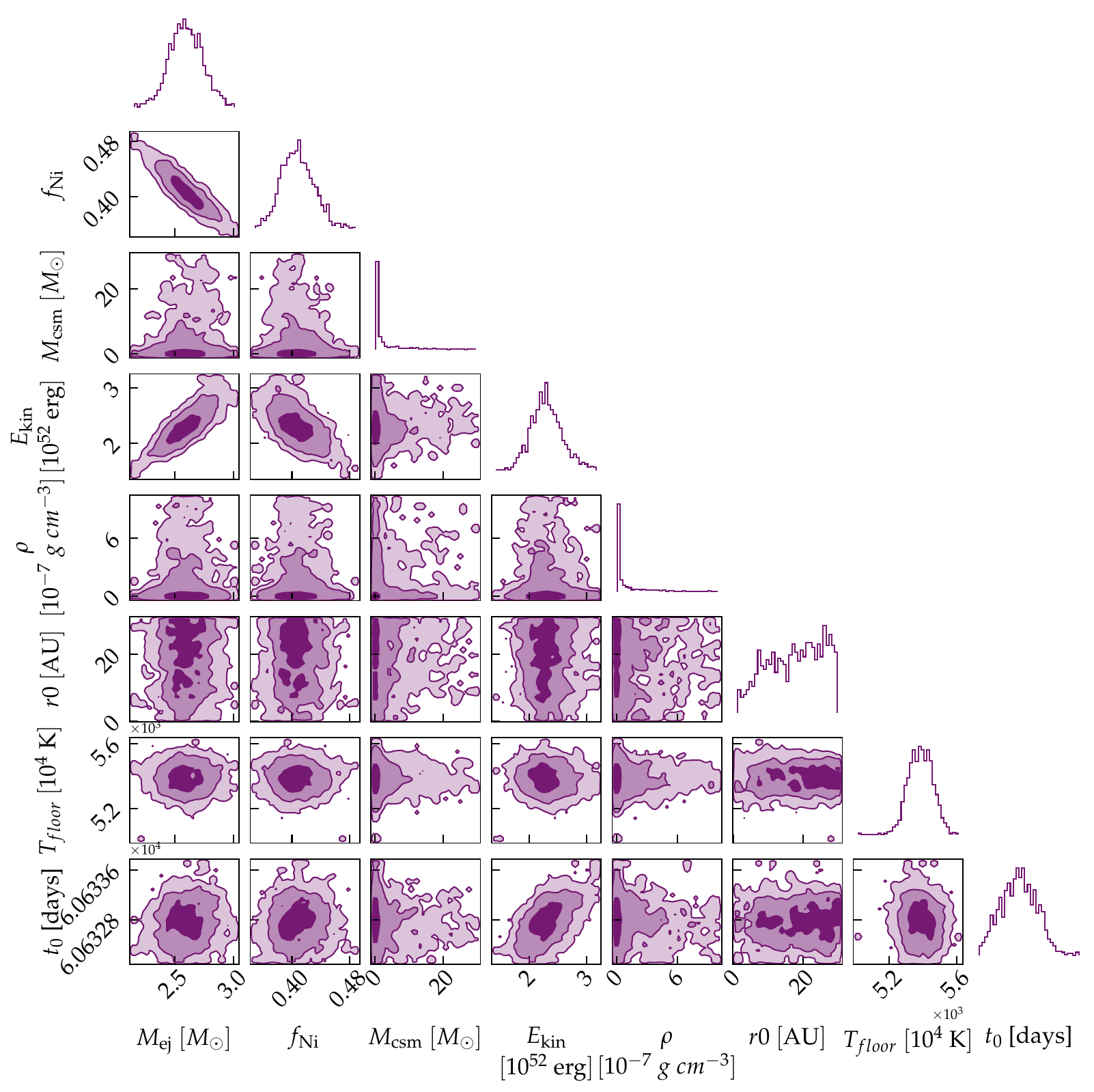}
  \end{minipage}
  \caption{\textit{Left:} The best-fit multi-band light curves of SN~2024abvb using the radioactive decay plus CSM interaction model \citep{Chatzopoulos2013,Villar2017}. The model exhibits difficulty reproducing the rise to peak in the IR bands and the post-peak decline in the UV bands. \textit{Right:} Corner plot of the model fit to the multi-band light curves of SN~2024abvb computed with \texttt{Redback}. See Table \ref{tab:csminteraction_table} for the best-fit parameters.}
  \label{fig:redbackcsm+nickel}
\end{figure*}

The light curves of normal Type Ib and Ic SNe are well-modelled by radioactive $^{56}$Ni decay-powered models (e.g., \citealp{Anderson2019}).  However, this is not always the case for interacting SESNe. The light curves of Type Icn SNe typically require a different and/or additional power source because $^{56}$Ni is inconsistent with the photometric properties post maximum light. Additionally, the presence of narrow-line emission in their spectra at early times (see Figure \ref{fig:spectra_timeseries}), rapid photometric evolution (see Figure \ref{fig:lightcurves}), high peak bolometric luminosities (see Figure \ref{fig:bbluminosity}), and blue colours (see Figure \ref{fig:colour_comparison}) suggests that CSM interaction is powering the light curve. 

However, to investigate whether the light curve is predominantly powered by radioactivity of $^{56}$Ni or by a combination of $^{56}$Ni decay and CSM interaction, we model the multiband UVOIR light curves of SN~2024abvb using \texttt{Redback}\footnote{https://redback.readthedocs.io/en/latest/index.html} \citep{Sarin2024}, a Bayesian inference package for modelling transients. We employ two models: the radioactive decay model from \citet{Arnett1982} (see \ref{radioactivedecaymodel}) and the radioactive decay model with additional CSM interaction from \citet{Villar2017} and \citet{Chatzopoulos2013}. The free parameters are inferred using nested sampling with \texttt{Nestle}\footnote{http://kylebarbary.com/nestle/} \citep{Barbary2021}, assuming a Gaussian likelihood.

The presence of narrow \CII\, and \OI\, P~Cygni features, together with the bolometric properties, blue colours, and rapid evolution of SN~2024abvb, indicates CSM interaction as the dominant power source. Radioactive decay likely provides an additional contribution, similar to the scenarios proposed by e.g. \cite{Perley2022, pellegrino_diverse_2022} and \cite{Davis2023}. We model this scenario by fitting the multiwavelength light curves with a combination of the Arnett model and the semi-analytical CSM interaction model of \citet{Chatzopoulos2012}, implemented as the \texttt{csm\_nickel} model in \texttt{Redback}. 
Similar to \citet{pellegrino_diverse_2022}, we fixed the optical (electron scattering) opacity to 
$\kappa_o = 0.04~\mathrm{cm}^{2}~\mathrm{g}^{-1}$, 
and gamma opacity to $\kappa_\gamma = 0.08~\mathrm{cm}^{2}~\mathrm{g}^{-1}$. The kinetic-to-thermal conversion efficiency, $\epsilon$, is fixed at 0.5 to enable direct comparisons with \citet{pellegrino_diverse_2022} and \citet{Davis2023}.
The $\mathrm{f}_{\mathrm{Ni}}$ ($\text{f}_{\text{Ni}}=\frac{\text{M}_{\text{Ni}}}{\text{M}_{\text{ej}}}$) priors were constrained initially to be $<0.008$ ($<0.004$~$\mathrm{M}_\odot$), which was taken from the published 
$\mathrm{M}_{\mathrm{Ni}}$ values for the SNe~Icn sample in \citet{pellegrino_diverse_2022}. 
We assumed that the best-fit value of $<1$\Msun~derived from the radioactive decay fit would be outside the fiducial range for CC SNe (see \ref{radioactivedecaymodel}). However, this constraint yielded no fits. Instead, we used broad priors for the nickel fraction, ranging from 0.0001 to 0.50.

Log-uniform priors are adopted for the ejecta mass and CSM mass, motivated by the reported values for the SNe~Icn population from \citet{pellegrino_diverse_2022} (see Table \ref{tab:csminteraction_table}). A log-uniform prior of $10^{-11} -10^{-6}$ g cm$^{-3}$ is assumed for the CSM density. The inner CSM radius is assigned a uniform prior over $0.1-30.0$ au, consistent with the range of radii determined by  \citet{pellegrino_diverse_2022}. 
The initial prior on $\eta$, the CSM density profile index, was chosen as a uniform distribution between $\eta=0$ (a uniform density CSM shell) and $\eta=2$ (CSM due to a steady mass-loss wind). The resulting posterior distribution reaches the upper bound of this prior, favouring a wind-like CSM structure for SN~2024abvb. However, to enable comparison with the literature, we fix $\eta$ to 2, as in e.g., \citet[][]{pellegrino_diverse_2022}. 

The best-fit light-curve and corner plots are shown in Figure \ref{fig:redbackcsm+nickel}, with parameters listed in Table \ref{tab:csminteraction_table}. The evolution of SN~2024abvb is consistent with an ejecta mass of $\sim$2.59\,\Msun\ interacting with a CSM mass of 0.28\,\Msun. Similar to what was found by \citet{Fraser2021} and \citet{pellegrino_diverse_2022}, this ejecta mass lies at the upper end of values inferred for other SNe~Icn (e.g., $\text{M}_{\text{ej}}\sim$2$\,\text{M}_{\odot}$ for SN~2021csp), yet remains below expectations for WR progenitors with pre-SN explosion masses $>10$\,\Msun. The inferred CSM mass is consistent with that found by \citet{Gangopadhyay2025B} ($\sim0.22$\,\Msun)  for SN~2023xgo and by \citet{pellegrino_diverse_2022} for SN~2019hgp ($\sim0.26$\,\Msun). However, the posterior for $\text{M}_{\text{CSM}}$ is weakly constrained compared to other parameters in our analysis (see Figure \ref{fig:redbackcsm+nickel}). Consequently, the quoted value should be treated with caution. The inferred $^{56}$Ni mass of $\sim0.1$\,\Msun\ exceeds that measured for other SNe~Icn by roughly one order of magnitude (e.g., \citealt{pellegrino_diverse_2022}), but is consistent with values reported for SNe~Ibn (e.g., \citealt{Pellegrino2024}). \citet{Maeda2022} found that SNe~Ibn produce very little $^{56}$Ni mass, with the upper limits $\sim0.1$\,\Msun, compared to canonical core-collapse supernovae (CCSNe). 

Our fit also yields a CSM mass of $\sim0.28$\,\Msun\ and a CSM radius of $\sim18.5$\,au. The inferred CSM mass lies within the range reported for SNe~Icn,  (0.12\,\Msun$<\rm{M}_{\rm{CSM}}<0.73$\,\Msun) whereas the CSM radius is among the highest in the SNe~Icn sample, which typically spans 0.05\,au\,$<R_{\rm{CSM}}<0.13$\,au \citep{pellegrino_diverse_2022}. \citet{Pellegrino2022} found that the light curves of four SNe~Ibn can be reproduced with models featuring ejecta masses of $\approx1-3$\,\Msun, CSM masses of $\approx0.2-1$\,\Msun, and CSM radii of $\approx20-65$\,au. However, their derived $\text{M}_{\text{Ni}}$ is comparable to those reported for SNe~Icn. This similarity suggests that SN~2024abvb and SNe~Ibn may share a common progenitor system (see Section \ref{IcnIbncomparison}). 

The best-fit light curve in Figure \ref{fig:redbackcsm+nickel} shows that the CSM+Ni model does not accurately reproduce the rise in the redder IR bands nor the post-peak decline in the UV bands. This suggests that the model may be insufficient to fully capture the behaviour of SN~2024abvb. In particular, the model assumes a uniform CSM density and spherical symmetry, whereas this object may instead feature a clumpy or asymmetric CSM. \citet{Gangopadhyay2025B} applied the same CSM+Ni model implemented in \texttt{Redback} for their transitional candidate, SN~2023xgo, finding ejecta masses, kinetic energy, and CSM radius that are substantially lower than those inferred for SN~2024abvb. They also report difficulty in reproducing the late-time light curve beyond $\sim30$ days post-maximum. Similarly, we find that the model struggles to fit the data at phases $\gtrsim 20$ days post-explosion, indicating that the inferred parameters may be sensitive to the adopted priors, even when these priors are consistent with those in the literature. In particular, the CSM density, radius, and mass are degenerate, so uncertainties in one parameter propagate directly to the others.

\begin{table}[]
    \centering
    \caption{Best-fit parameters for the Arnett Law plus CSI model. The light curves are well reproduced with a considerably large ejecta mass compared to the SNe~Icn and Ibn sample.}\label{tab:csminteraction_table}
    \begin{tabular}{lll}
        \hline
        Parameters & Priors &  Best-Fitted Values  \\
        \hline \hline
        \\    
           
            $\text{M}_{\text{ej}}~[\text{M}_{\odot}]$ & log$~\mathcal{U}~[0.1, 30] $  & $2.59_{-0.13}^{+0.14}$         \vspace{0.50mm} \\
            $\text{f}_{\text{Ni}}$ &  $\mathcal{U}~[0.0001, 0.50]$    & $0.41_{-0.02}^{+0.02}$      \vspace{0.50mm} \\
            $\text{M}_{\text{csm}}~[\text{M}_{\odot}]$ & log$~\mathcal{U}~[0.01, 30] $  & $0.28_{-0.25}^{+4.94}$         \vspace{0.50mm} \\
            $\text{E}_{\text{kin}}~[\text{erg}]$ & log$~\mathcal{U}~[5\times10^{49}, 5\times10^{52}]$  & $2.29_{-0.24}^{+0.28}\times10^{52}$      \vspace{0.50mm} \\
            $\rho~[\text{g~cm}^{-3}]$ & log$~\mathcal{U}~[10^{-11}, 10^{-6}]$  & $0.14_{-0.14}^{+2.56}\times10^{-10}$      \vspace{0.50mm} \\
            $\text{r}_{\text{0}}~[\text{AU}]$ & $\mathcal{U}~[1.0, 30]$  & $18.50_{-9.91}^{+7.99}$      \vspace{0.50mm} \\
            $\text{T}_{\text{floor}}~[10^{4}\text{K}]$ &  log$~\mathcal{U}~[2\times10^{3}, 10^{4}]$ & $5386.92_{-73.89}^{+70.59}$ \vspace{0.50mm}\\
            $\text{t}_{\text{0}}~[\text{days}]$ &  $\mathcal{U}~[t_{exp}-2, t_{exp}+2]$ & $60632.77_{-0.29}^{+0.29}$\\
 \vspace{0.20mm} \\
            \hline \\
    \end{tabular}
\end{table}

\subsection{Spectroscopic Properties and evolution}\label{sec:spec_properties}

\begin{figure*}[!htb]
    \centering
    \includegraphics[width=\textwidth]{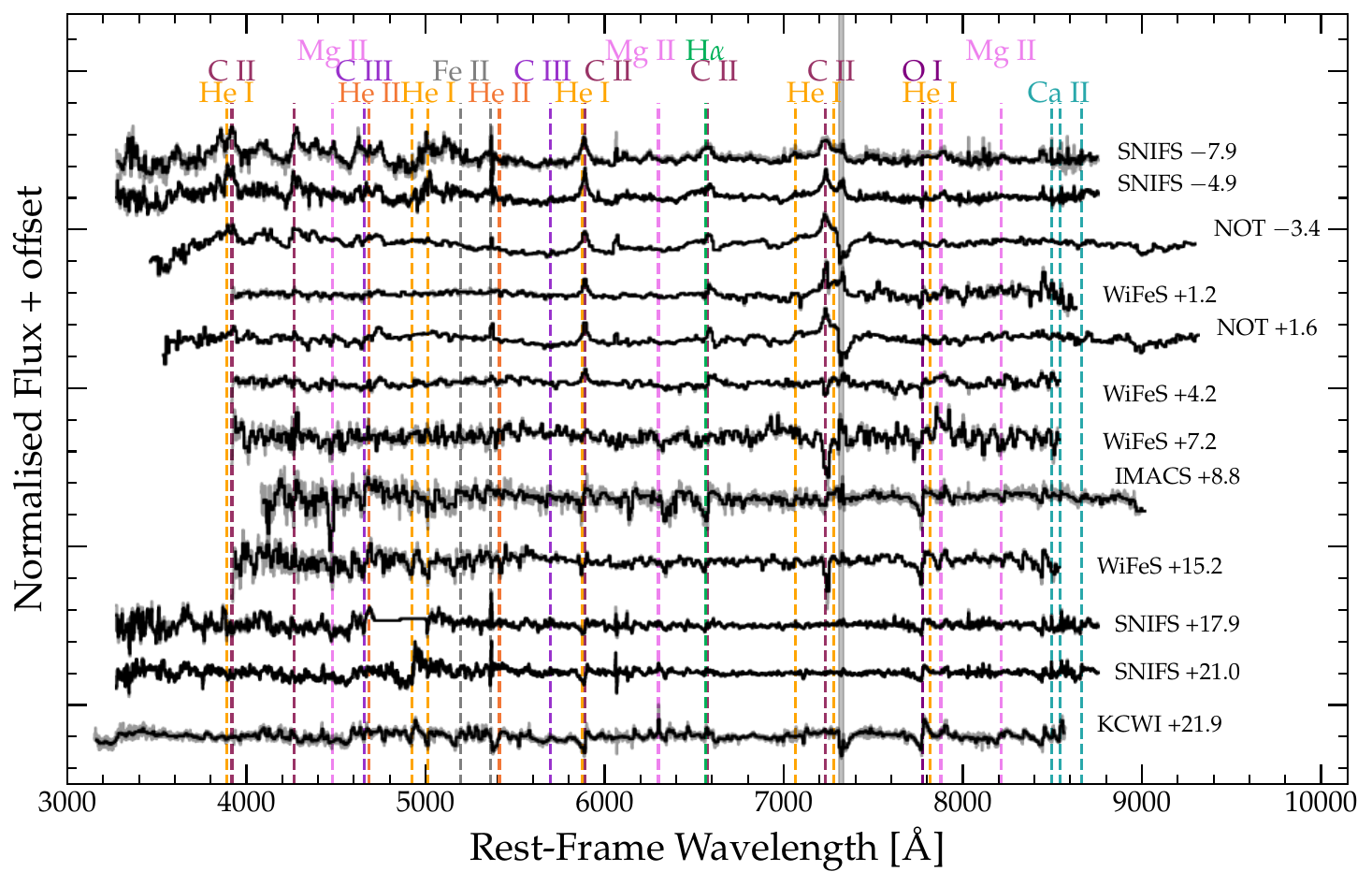}
    \vspace{-2mm}
    \caption{The optical spectral timeseries of SN~2024abvb. All fluxes have been normalised and vertically offset for clarity. The data has been binned into $5\,$\AA-wide bins from the original observations. Phases are calculated relative to the $o$-band maximum light on MJD 60645.27. Dashed lines correspond to prominent emission features. The grey shaded region indicates telluric absorption bands. }\label{fig:spectra_timeseries}
    \vspace{-2mm}
\end{figure*}

From the ANU 2.3m WiFeS observation taken +1.2 days after peak brightness, a spectrum of the transient and its host galaxy were extracted. As there is no previously known redshift for the host galaxy, the redshift of SN~2024abvb was measured using the automatic redshift software \texttt{MARZ}\footnote{https://samreay.github.io/Marz/} \citep{Hinton2016} that matches the input spectrum against galaxy templates. By aligning the narrow \Ha\, in the host-galaxy spectrum, a redshift of $z=0.0391 \pm 0.0002$ was determined. This value is consistent with the redshift provided by the TNS classification of $z=0.039$\footnote{https://www.wis-tns.org/object/2024abvb}. 

The optical spectral sequence of SN~2024abvb is presented in Figure \ref{fig:spectra_timeseries}. Table \ref{tab:spectra_table} presents a log of the spectra. The earliest spectrum of SN~2024abvb at -7.9~d exhibits narrow emission features arising from C, He (and potentially H) that originate from the recombination of the surrounding CSM. These notable features are identified at $3910~\text{\AA}$, $4270~\text{\AA}$, $5890~\text{\AA}$, $6560~\text{\AA}$, and $7230~\text{\AA}$. The first peak is due to a blend of \CII~$\lambda3915$ and \HeI~$\lambda3888$, and the second peak at $4267~\text{\AA}$ is due to \CII. The third and fifth peaks are a combination of \CII~ and \HeI~ lines; the latter peak has a blue wing, while the former is symmetrical. The fourth peak is dominated by \CII\, emission, with likely weak \Ha\, contribution due to the proximity of the neighbouring \Ha\, line. Overall, the narrow-line features are seen until 7.2~days after maximum light. 

In addition, there is a weak \OI~$\lambda7774$ P~Cygni line that gradually strengthens $\sim7-22$ days after maximum light. Additionally, we see emission signatures from the \MgII~$\lambda$4481, $\lambda$7877, $\lambda$7896 lines, and weak \CaII~triplet lines. The \MgII~$\lambda$4481 emission line is also observed in some SNe~Ibn (SNe~Icn) such as SN~2006jc \citep{Pastorello2007} and SN~2019hgp \citep{pellegrino_diverse_2022}. The early presence of Mg lines suggests that the outer H and He shells of the progenitor have been stripped via stellar winds or binary interaction, which exposes the deeper O-Mg layer \citep{Kuncarayakti2022}.

Several \He features are present in the first two spectra of SN~2024abvb (-7.9~d and -4.9~d prior to peak brightness). In the forest of highly ionised lines blueward of $\sim5400~\text{\AA}$, narrow \HeI\, emission lines at $\lambda4922$ and $\lambda5015$ fade by the third spectrum (-3.4~d). It is difficult to discern whether \HeI\, $\lambda5876$ is a strong feature since it is coincident with the strong \CII\, $\lambda5890$ line. \HeI\, $\lambda7065$ and $\lambda7281$ emission is also identified on the blue and red wings of \CII\, $\lambda7234$ during the first four spectra, but disappears by +1.6 days after peak. 

From +4.2 days until +15.2 days, the \CII\, $\lambda7234$ line has evolved to an absorption line. However, this absorption feature disappears at +17.9 days. 
Between +4.2 days and +7.2 days, the narrow \HeI\, and \HeII\,  emission transitions to absorption features. At the latter phase, \HeI\, $\lambda4922$, $\lambda5015$ absorption lines appear and continue to strengthen until +21.9 days. From 15.2 days after peak brightness, we see a gradual strengthening of \HeI\, $\lambda 5875$ and \OI\, $\lambda 7774$ P~Cygni lines.

\begin{figure*}[hbt!]
    \centering
    \includegraphics[width=\columnwidth]{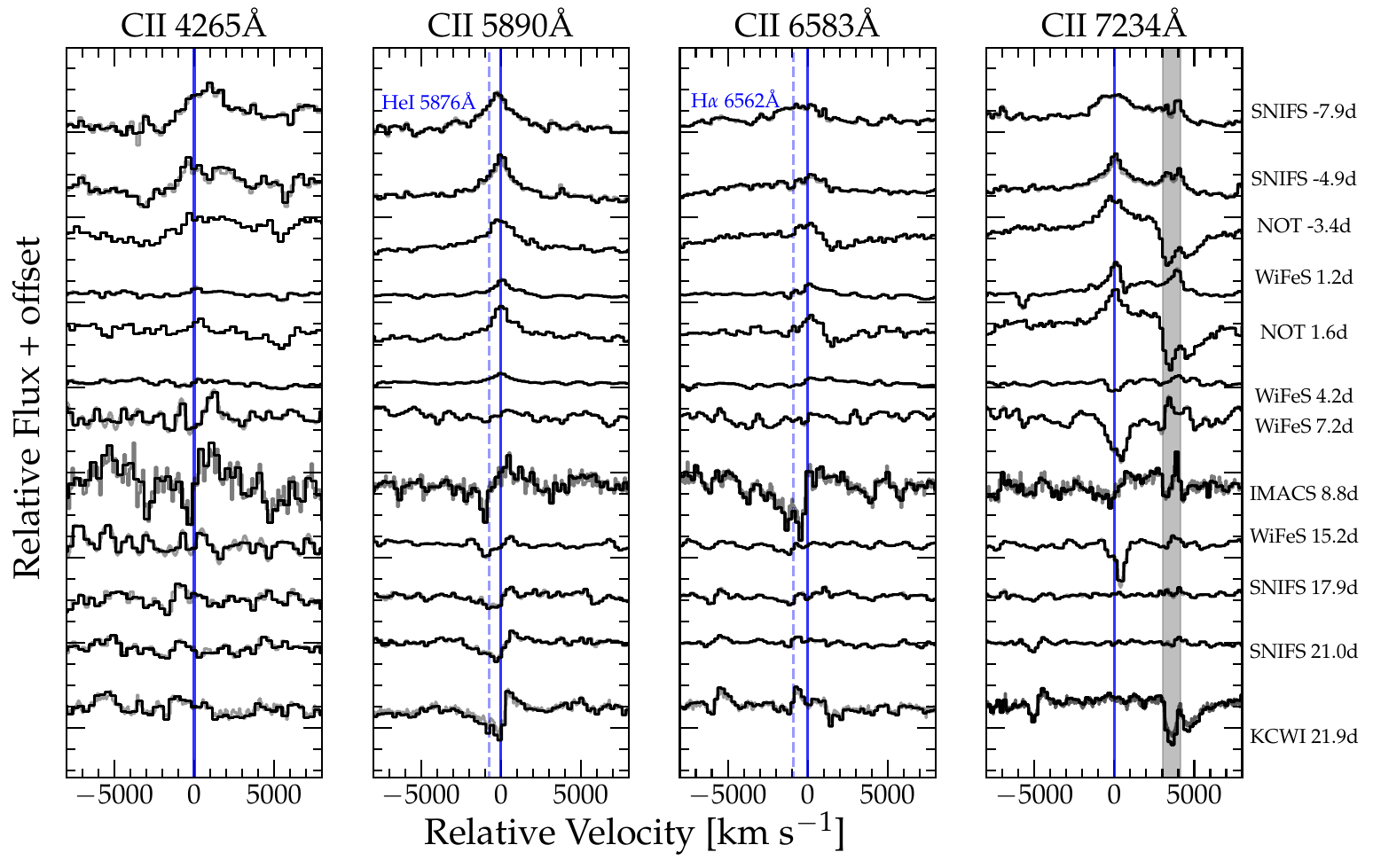}
    \vspace{-2mm}
    \caption{Evolution of the \CII\, features in 2024abvb. We plot $4265\,\text{\AA}$, $5890\,\text{\AA}$, $6583\,\text{\AA}$, and $7234\,\text{\AA}$ on a velocity scale relative to their respective rest wavelengths marked by blue solid lines. The blue dashed lines show neighbouring features that blend with the main line profile. Continuum-subtracted spectra plotted in black and binned into 5\,\AA-wide bins, while the light grey spectra are the original data. The phases relative to the $o$-band maximum light and the respective instrument are shown on the right.}\label{fig:CII_timeseries}
    \vspace{-2mm}
\end{figure*}

Figure \ref{fig:CII_timeseries} shows the isolation of \CII; $\lambda4265$, $\lambda5890$, $\lambda6583$, and $\lambda7234$. The early evolution of \CII\,$\lambda5890$ and \CII~$\lambda7234$ exhibits roughly symmetrical line profiles. At the time of strongest emission, the velocity for each line varies from about 2200 \kms to 900 \kms. At +8.8 days, absorption features emerge, with \CII\,$\lambda5890$ evolving into a P~Cygni feature. The \CII\,$\lambda5890$ and \CII\,$\lambda6583$ lines are likely blended with \HeI\,$\lambda5876$ and \Ha\,$\lambda6562$, respectively, which may alter the line profile. The spectral evolution of \CII\,$\lambda5890$ and \CII\,$\lambda6583$ is similar to that of SN~2022ann at comparable epochs (see Figure 9 in \citealt{Davis2023}).

\begin{figure}[hbt!]
    \centering
    \includegraphics[width=\columnwidth]{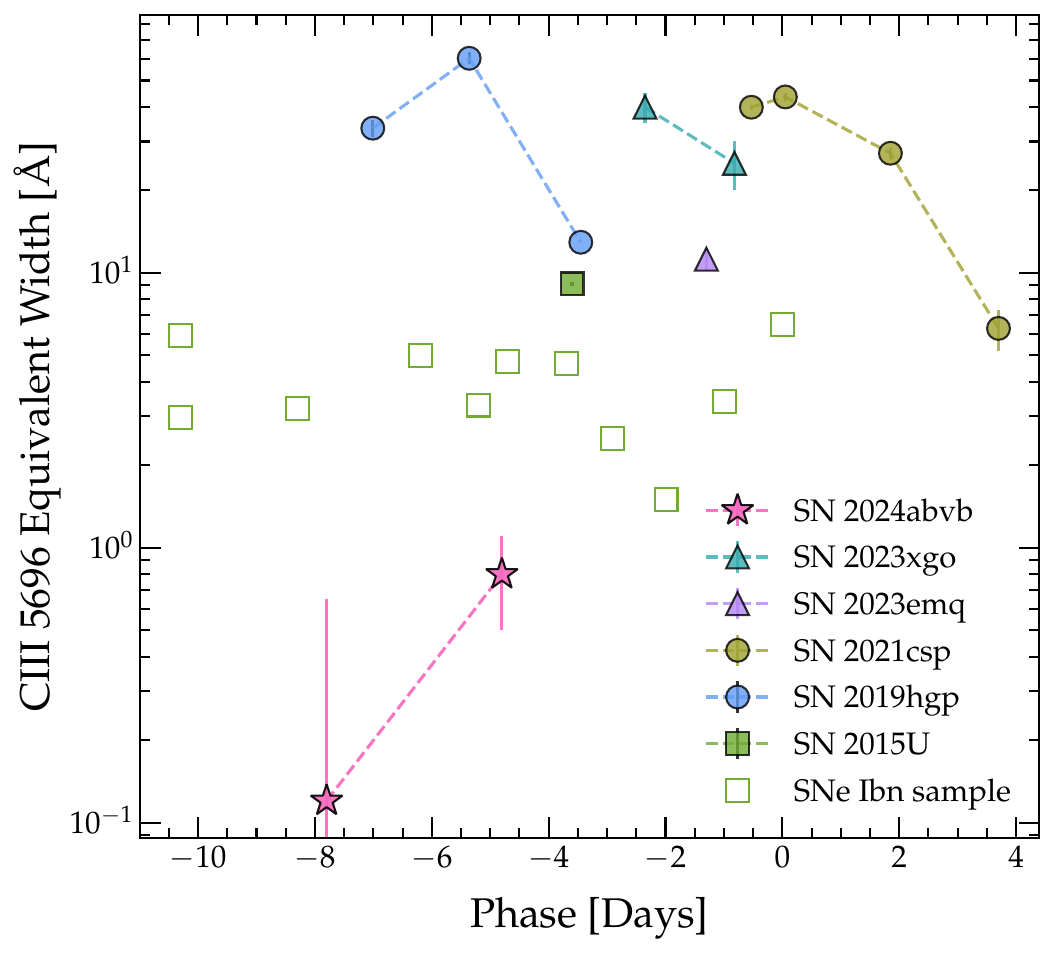}
    \vspace{-2mm}
    \caption{Pseudo equivalent width measurements of the \CIII\, $5700$\AA\, emission for SN~2024abvb (pink star), SNe~Icn 2021csp and 2019hgp (green and blue circles), SNe~Ibn/Icn candidates 2023emq and 2023xgo (purple and teal triangles), and flash-ionised SNe~Ibn (green squares). Open green squares indicate $3\sigma$ upper limits for SNe~Ibn with ambiguous \CIII\, detections (see Figure 5 from \citealt{Pursiainen2023}). Phases are relative to the $r$-band maximum light. SN~2024abvb does not resemble any comparison sample.}\label{fig:EW}
    \vspace{-2mm}
\end{figure}

In Figure \ref{fig:EW} we show the pseudo-equivalent widths (pEWs) of the \CIII\, emission at 5700\AA~ for the early spectra of SN~Ibn, SN~Icn, and transitional SN~Ibn/Icn candidates compared with SN~2024abvb. The \CIII~5700\AA~ emission in SN~2024abvb is the weakest among the sample; with estimated pEW values of $0.12\pm0.53\,\text{\AA}$ and $0.80\pm0.30\,\text{\AA}$ at $-7.9$ d and $-4.9$ d relative torr peak brightness, respectively. These values are approximately an order of magnitude smaller than those measured for SNe~Ibn, and about two orders of magnitude smaller than the pEWs observed in SNe~Icn. The weakness of \CIII~suggests that the emitting region lacks carbon and has low ionisation conditions. 

\subsubsection{Pre-Peak Optical Spectral Evolution}\label{pre-peak}
We compare the spectra of SN~2024abvb with a sample of SNe~Ibn, Icn, and transitional Ibn/Icn, which show signatures of narrow-line features in their early spectra. For the SNe~Ibn (blue spectra), the following objects were selected: PS1-12sk \citep{Sanders2013} and SN~2024acyl \citep{Dong2025}, the only SNe~Ibn with projected distances larger than that of SN~2024abvb (see Section \ref{sec:host-galaxy}); SN~2022pda \citep{Fulton2022}, which has an early pre-peak spectrum obtained at a phase similar to the first spectrum of SN~2024abvb. The sample also includes SN~2023ubp \citep{Tonry2023, Hinkle2023}, SN~2022ablq \citep{Pellegrino2024}, SN~2023tsz \citep{Warwick2025}; which exhibit \CII\, emission similar to that of SN~2024abvb at comparable phases, and SN~2006jc \citep{Pastorello2007}; the prototypical SN~Ibn. Among the SNe~Icn (green spectra), we compared SN~2024abvb with SN~2019hgp, SN~2019jc, SN~2021csp, SN~2021ckj, and SN~2022ann. We also included the only two transitional Ibn/Icn SNe (pink spectra) found in the literature: SN~2023emq and SN~2023xgo. 
All the comparison spectra were downloaded from WISeREP \citep{Yaron2012} and corrected for the host-galaxy redshift.

\begin{figure*}[p]
\centering
    \includegraphics[width=0.86\linewidth]{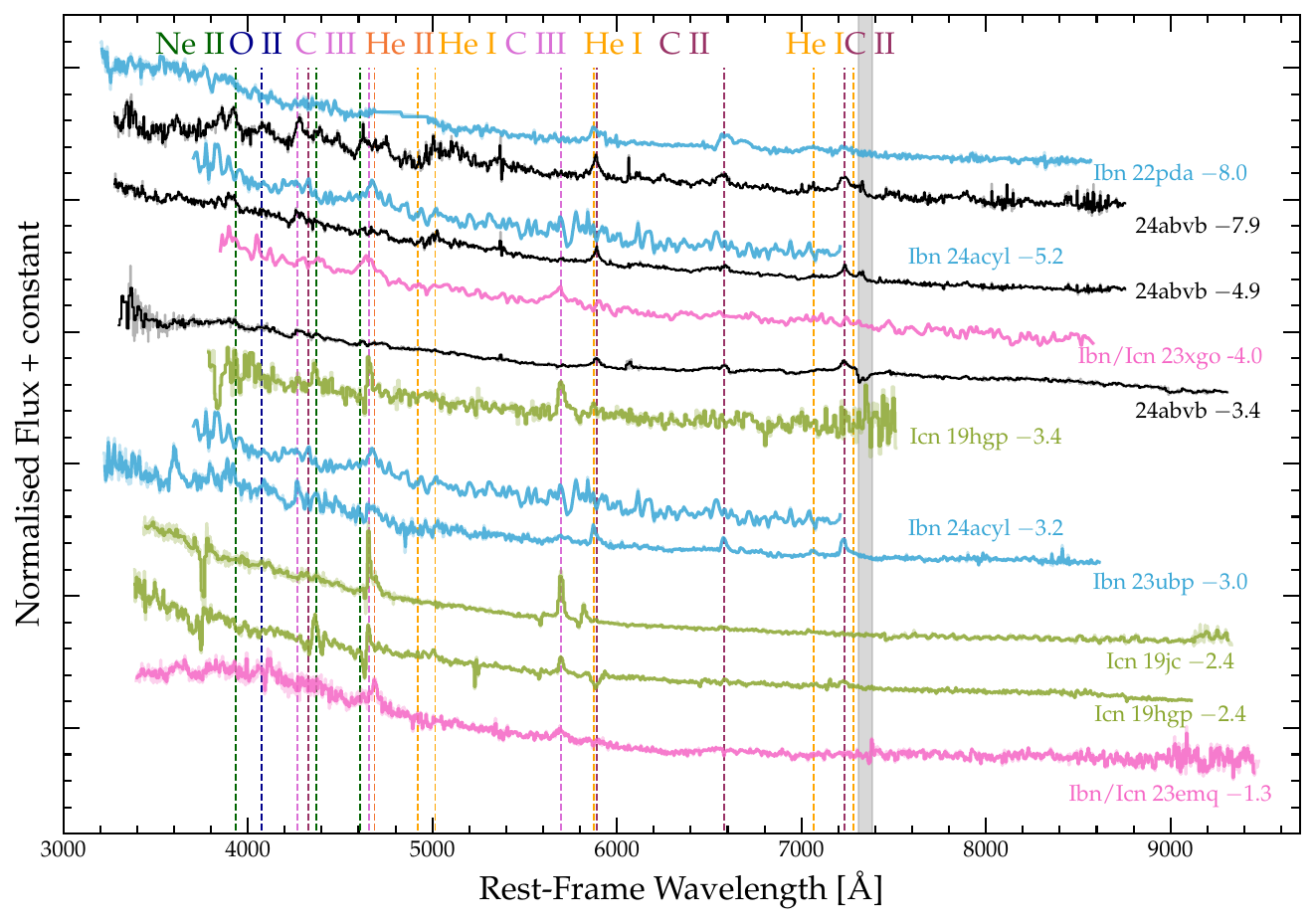}
    \vspace{-2mm}
    \caption{Comparison of SN~2024abvb spectra with Type Ibn (blue), Icn (green), and Ibn/Icn SNe events (pink) at pre-peak epochs. The prominent spectral lines are marked, and the indicated phases are relative to the maximum light. All spectra are binned into 5\,\AA-wide bins from the original data. The grey shaded regions indicate telluric absorption bands.}\label{fig:spectra_prepeak_compare}

\end{figure*}

In Figure \ref{fig:spectra_prepeak_compare}, we compare the early-time optical spectra of SN~2024abvb with the aforementioned SNe sample. The first spectrum of SN~2024abvb was obtained $\sim$8 days before maximum, making it one of the earliest epochs in our sample of interacting SNe, second to only SN~2022pda. Compared to SN~2022pda, both SNe show strong \CII\, lines and exhibit a hot blue continuum. All the SNe~Icn spectra exhibit \CIII\, emission features at $4658\,\text{\AA}$ and $5696\,\text{\AA}$, which are typical features seen in the SNe~Icn sample in \citet{pellegrino_diverse_2022}. The $5696\,\text{\AA}$ emission is present in SN~2023emq but is weaker than the SNe~Icn, SN~2019hgp, and SN~2019jc. These emission features are extremely faint in SN~2024abvb. Instead, we see strong peaks of \CII\, blended with \HeI\, at $5890\,\text{\AA}$ and $7230\,\text{\AA}$, which are two of the most prominent emission lines in SN~2024abvb that are not observed (or rather very faint as for SN~2019hgp at -3.4 days (5890\,\AA) and SN~2019hgp (7230\,\AA)) in the SNe~Icn sample. 

A notable feature seen in the very early spectra of SN~2024abvb is the lack of \CIII\, emission at $\sim 5700\,\text{\AA}$ compared to other SNe. \CIII\, $5696\,\text{\AA}$ is the most prominent in SNe~Icn and SNe~Ibn/Icn, but is not seen in SN~2024abvb and SNe~Ibn: SN~2022pda and SN~2023ubp. SN~2024abvb also shares similar \CII\, emission features with our SNe~Ibn sample. Both SNe~Ibn spectra contain strong \CII\, emission features at $5890~\text{\AA}$, $6583~\text{\AA}$, and $7234~\text{\AA}$, which is also seen in SN~2024abvb. 

Ne is rarely seen in CCSNe spectra \citep{pellegrino_diverse_2022}. Yet in Figure \ref{fig:spectra_prepeak_compare}, we see \NeII\, features in the early time spectra of SN~2024abvb and Icns SN~2019jc and SN~2019hgp \citep{Gal-Yam2022}. In particular, the first two spectra of SN~2024abvb display strong \NeII\,$\lambda$3933 and $\lambda$4607 features. \citet{pellegrino_diverse_2022} suggest that this Ne originates from C-burning and was stripped from the inner layers of the progenitor. As such, measuring the Ne abundance in the CSM could help constrain progenitor channels. 

\begin{figure*}[p]
    \includegraphics[width=0.86\linewidth]{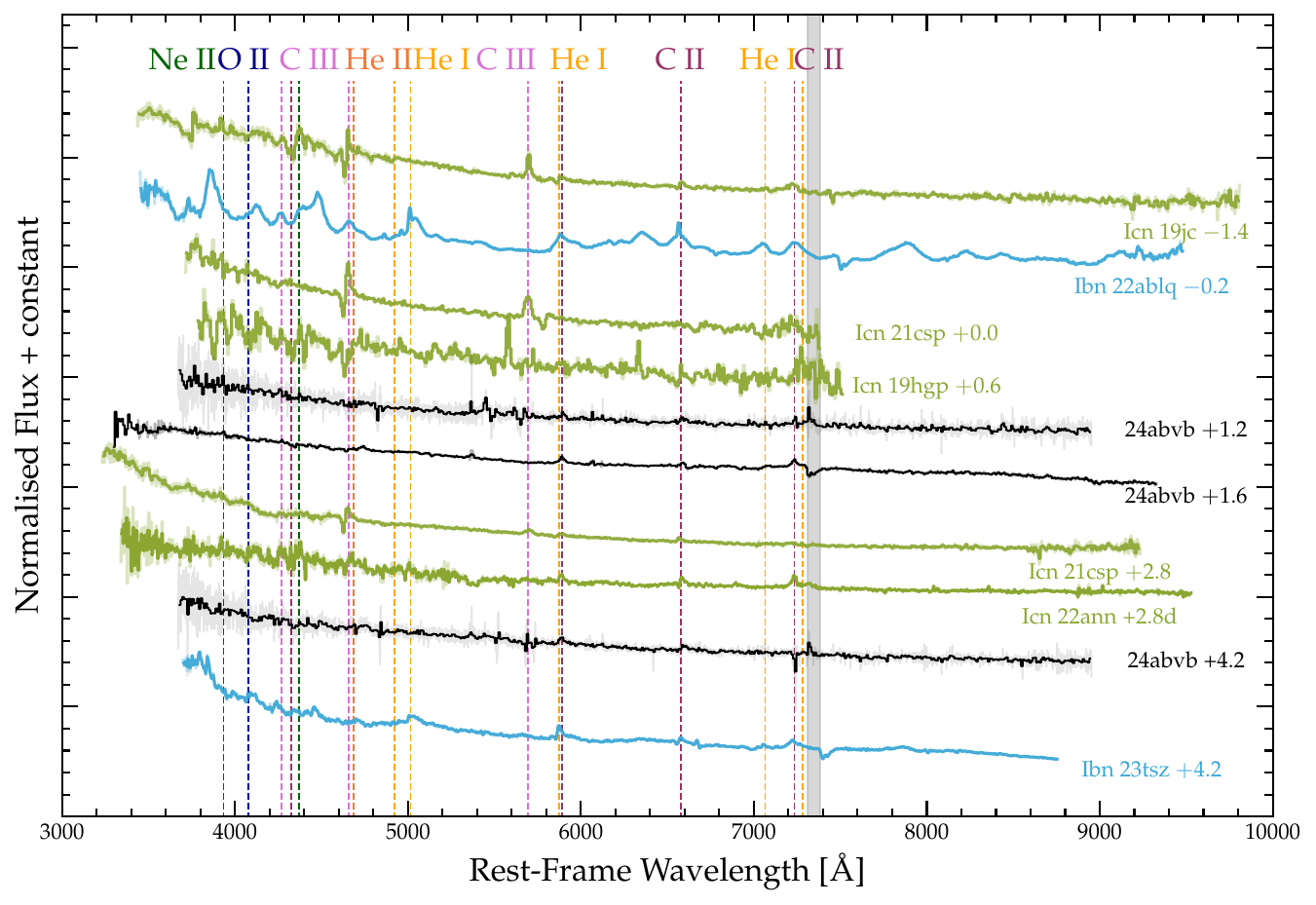}
    \vspace{-2mm}
    \caption{The spectral evolution of SN~2024abvb near peak-brightness ($-1.4$ to $+5$ days) compared with SNe Icn (green) and Ibn (blue). SN~2024abvb shows \CII\, features similar to SN~2022ann and SN~2023tsz.}\label{fig:spectra_peak_compare}

\end{figure*}

The spectroscopic sequence at peak brightness (from -1.4 to +5 days) in Figure \ref{fig:spectra_peak_compare}, shows that SN~2024abvb is highly similar to SNe~Ibn 2023tsz and SN~Icn 2022ann. SN~2024abvb, like other SNe~Icn lacks \HeI and \HeII\, features. Unlike some SNe~Icn, it shows no early significant P~Cygni absorption blueward of $5000$\AA\, or \CIII\, emission (see \citealt{pellegrino_diverse_2022}). Instead, \CII\, features dominate, notably a strong 7234\,\AA\ and a weaker $6583~\text{\AA}$, line, similar to SN~Icn 2022ann. SN~2022ann also does not exhibit strong \CIII\, lines which are seen in SN~Icn 2021csp or 2019jc. SN~2024abvb does not show any broad \HeI\, and \HeII\, emission lines like in SNe~Ibn 2022ablq, but rather weak \He\,lines like in SN~2023tsz. However, the \CII\, features present in these SNe are alike. In this case, SN~2024abvb has features most consistent with SN~2023tsz from the SNe~Ibn sample and SN~2022ann from the SNe~Icn sample.

\subsubsection{Post-Peak Optical Spectral Evolution}
\begin{figure*}[h!]
    \includegraphics[width=\linewidth]{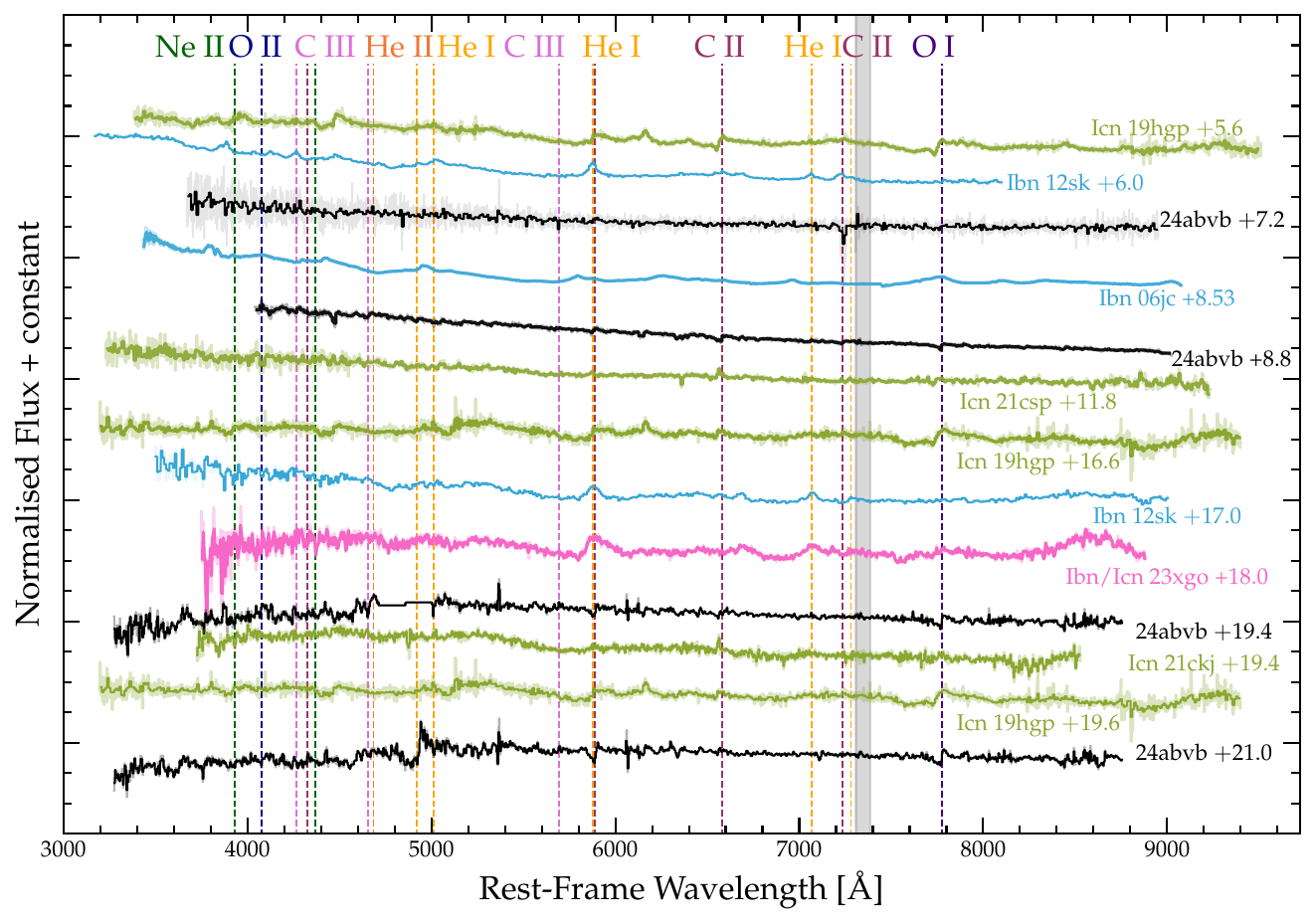}
   
    \caption{Post-peak spectral evolution of SN~2024abvb compared with Type Ibn (blue), Icn (green), and transitional Ibn/Icn (pink) events. Unlike the SNe~Ibn, SN~2024abvb shows no broad \He emission at this phase, instead exhibiting strong \CII\, and \OI\, P~Cygni features similar to the Type Icn SN~2019hgp.}\label{fig:spectra_postpeak_compare}
    \vspace{-2mm}
\end{figure*}

In Figure \ref{fig:spectra_postpeak_compare}, the spectral behaviour of SN~2024abvb diverges from the typical behaviour of SNe~Ibn at late times. From +7.2\, days post peak, the strong \CII\, features have faded and P~Cygni features have developed in SN~2024abvb. Apart from the \CII\, P~Cygni features (see Figure \ref{fig:CII_timeseries}) we also see a \OI\, 7774\,\AA\, P-Cygni line emerging, which is consistent with SNe~Icn \citep{pellegrino_diverse_2022}. At this epoch, SN~2024abvb more closely resembles SNe~Icn 2019hgp and 2021csp. For the SNe~Ibn (SN~2006jc and PS1-12sk), broad intermediate-width lines can be seen at \HeI\, $5876$\,\AA\, and \OI\, 7774\,\AA\,. However, we do not observe broad \He\, emission lines in SN~2024abvb but rather small absorption dips at $4922$\,\AA\, and $5015$\,\AA. 
Starting around 19 days after peak brightness, the strongest features in SN~2024abvb are the \OI\, and \CII\, P~Cygni lines, coinciding with the onset of light-curve decline. At a similar phase, SN~2022ann \citep{Davis2023} also exhibits these P~Cygni features. However, by about 60 days after peak, they transition to forbidden emission lines, indicating a shift to the nebular phase. Our dataset is limited to the photospheric phase, so we are unable to detect potential forbidden emission lines. 

\citet{pellegrino_diverse_2022}, and \citet{Perley2022} report a transition from a narrow-line phase to a broad-line phase in their spectra. For example, \citet{pellegrino_diverse_2022} highlight that around 3 weeks after maximum light, broad spectral features (with velocities of $\sim 10,000$\kms) start to emerge (e.g., a broad \CaII\, NIR triplet in SN~2021csp). On the other hand, SN~2019hgp maintains its narrow features for longer than SN~2021csp but eventually develops broad P~Cygni features. \citet{Gangopadhyay2025B} also identified this transition in SN~2023xgo, but for the \HeI\, $5876$\,\AA\, line and suggest the broad features are a signature of the SN ejecta while the narrow features arise from the undisturbed He-rich CSM. However, like SN~2022ann, such a transition is not present in SN~2024abvb; the narrow \CaII\, NIR triplet persists and only slightly strengthens $\sim15$ days after peak. We were unable to obtain subsequent optical spectroscopy to search further for broad lines because SN~2024abvb became too faint before such lines emerged.


\subsubsection{Near Infrared Spectrum}
\begin{figure}
    \centering
    \includegraphics[width=\linewidth]{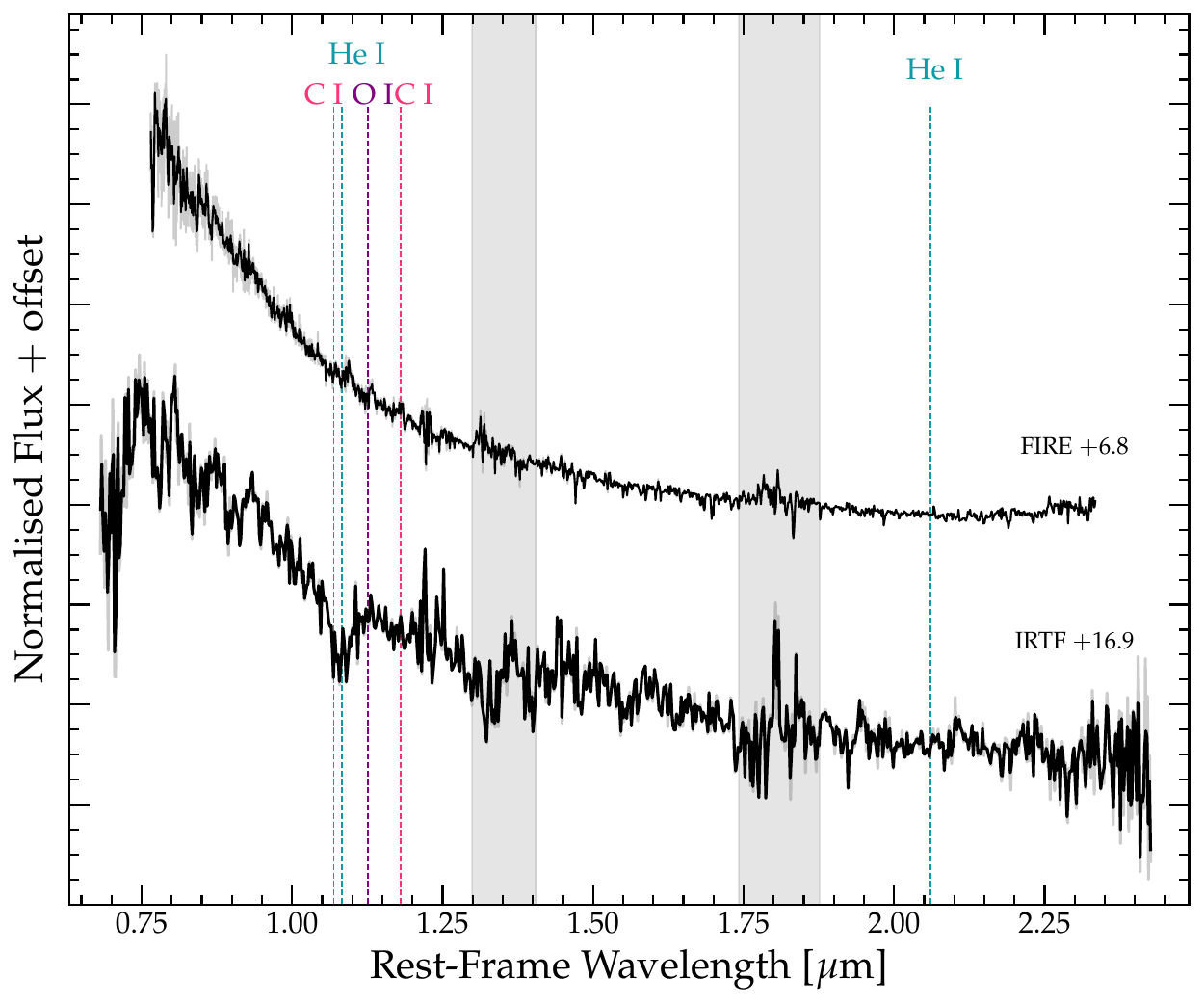}
    \vspace{-2mm}
    \caption{NIR spectra for SN~2024abvb. Each spectrum is binned from the original data into $10$\,\AA-wide bins. Phases are relative to $o$-band maximum light. Dashed lines correspond to spectral features seen in other SNe~Icn NIR spectra. Grey shading denotes the telluric regions.}\label{fig:NIR_spectrum}
    \vspace{-2mm}
\end{figure}

Figure \ref{fig:NIR_spectrum} presents the NIR spectra. One week after maximum light, the SED is dominated by a strong blue continuum. In the spectrum taken at +6.8 days, searches for \HeI, \CI, and \OI\ NIR features commonly seen in SESNe \citep{Shahbandeh2022} reveal no strong 1.08 $\micro\text{m}$ \HeI\ line (seen in 2023emq; \citealt{Pursiainen2023} and SN~2022ann \citealt{Davis2023}) nor the 1.13 $\micro\text{m}$ \OI\ line. We identify tentative 1.07 and 1.18 $\micro\text{m}$ \CI\ lines. In the spectrum taken at +16.9 days, weak 1.08 $\micro\text{m}$ \HeI\, 1.07 $\micro\text{m}$, and 1.18 $\micro\text{m}$ \CI\ lines are present. The contemporaneous optical spectrum ($+17.9$ days; SNIFS) also exhibits weak \He\ lines. At a similar epoch, strong 1.06 $\micro\text{m}$ \CI\ emission was detected in the Type Icn SN~2021csp \citep{Fraser2021}.  

\section{Discussion} \label{sec:discussion}
Our photometric and spectroscopic analysis indicates that SN~2024abvb is not strictly a Type Icn SN. We now synthesise the evidence to evaluate whether it more closely resembles a Type Ibn or Type Icn event. In this section, we also estimate the local SFR properties and present possible progenitor channels that could potentially explain the unique characteristics of SN~2024abvb.

\subsection{Host-Galaxy Offset}
One of the most striking characteristics of SN~2024abvb is its projected offset from its host galaxy. In Section \ref{sec:host-galaxy}, we found that SN~2024abvb has the largest offset (22.4 kpc) from its host-galaxy compared to other SNe~Icn in our sample (and the third largest among the SNe~Ibn sample). However, the question arises as to how SN~2024abvb arrived at its explosion site, assuming its progenitor originated from its host galaxy. 

Explaining the large offset of SN~2024abvb remains challenging. As a quick estimate, if the progenitor of SN~2024abvb was a single star with a zero-age main-sequence (ZAMS) mass of $M_\text{ZAMS}= 8\,$\Msun\, and a lifetime of $\approx50$\,Myr (see Figure 3. in \citealt{Zapartas2017}), it would require a velocity of $\sim$780\,\kms to move 22.4\,kpc during that time. On the other hand, if we assume the progenitor of SN~2024abvb is towards the higher mass limit, a $M_\text{ZAMS}= 40\,$\Msun\, star with a lifetime of $\approx7$\,Myr, this would require a velocity of $\sim3100$\,\kms. Such an extreme velocity is highly unlikely, given that it exceeds the velocity achievable through the Hills mechanism, and the fastest observed unbound main-sequence stars (hypervelocity stars) reach only $\sim1000$\,\kms~ \citep{Brown2015, Kreuzer2020}. These considerations suggest that the progenitor would more likely be a lower mass progenitor. A binary progenitor scenario is appealing because it allows for a lower-mass star; however, achieving near $1000$\kms\, in a binary system is difficult.

For the same upper mass-limit assumption ($M_\text{ZAMS}= 40\,$\Msun), \citet{Hosseinzadeh2019} estimated that SN~2006jc, the only SN~Ibn with a direct progenitor detection during an outburst, would require a velocity of $>$530\,\kms\, to reach its terminal location, substantially lower than the velocity required for SN~2024abvb. SN~2019jc has a projected offset of 11.2\,kpc which is the second largest in the SNe~Icn sample. \citet{pellegrino_diverse_2022} conservatively estimated that a runaway velocity of $\approx250$\,\kms, which is of the same order of magnitude as our estimate for a $M_\text{ZAMS}= 8$~\Msun\, progenitor, would be required for SN~2019jc to reach its explosion site within its lifetime. They further noted that if SN~2019jc originated from a WR star in a binary system, such a velocity is an order of magnitude lower than typical kick velocities imparted by companion SNe \citep{Renzo2019}, and that most massive stars at most travel hundreds of parsecs before exploding \citep{Eldridge2011}. Although, \citet{pellegrino_diverse_2022} suggested that such runaway velocities could be achieved through interactions with stellar clusters, the location of SN~2024abvb is devoid of star formation (see Section \ref{localenv}). We therefore tentatively rule out a (high-mass) WR star progenitor for SN~2024abvb. It is also important to note that the measured offset is a projected distance, so the true physical separation is likely larger, making these velocity estimates lower limits.

\subsection{SN~2024abvb: Ibn or Icn?}\label{IcnIbncomparison}
\subsubsection{Considerations for a SN~Ibn}

Even though SN~2024abvb was spectroscopically classified as a Type Icn after peak, it exhibits photometric and spectroscopic properties similar to those of SNe~Ibn. For example, the bolometric luminosity evolution and peak $r$-band magnitude match the SNe~Ibn template from \citet{Hosseinzadeh2017} (see Figure \ref{fig:bbluminosity} upper panel, grey band). Turning to the spectra, the early phase of SN~2024abvb could lead to a classification as a SN~Ibn due to the abundance of \CII~ lines and lack of prominent \CIII~ features, although early emission of \CIII~ $\lambda5700$ has been observed in SNe~Ibn (e.g., SN~2023emq \citep{Pursiainen2023} and 7 others in the SNe~Ibn sample from \citet{Farias2025}). While the presence of C emission does not determine whether a SN is classified as a SN~Ibn or SN~Icn, it is interesting to note that early \CII~ emission is generally more prevalent in SNe~Ibn, whereas \CIII~ is more common in SNe~Icn. Generally, the detection of \CII~ and \CIII~ lines suggests C-rich outer layers in the progenitor star. Since \CIII~ requires a higher ionisation potential than \CII, its presence implies hotter ejecta, whereas \CII~ emission may indicate carbon-rich, and potentially hydrogen-rich, CSM, possibly arising from episodic mass loss prior to explosion. By day 4 post-explosion, SN~2024abvb is not dominated by strong narrow \He\, lines, unlike the SNe~Ibn at similar phases. As such, based on its early-time spectra ($-7.9$ to $-3.4$ days relative to peak light) SN~2024abvb shows similarities to SNe~Ibn. 

The inferred blackbody and CSM properties of SN~2024abvb are broadly consistent with those of the SNe~Ibn population. Its initial blackbody temperature, for instance, closely matches those of SNe~Ibn (see the bottom panel of Figure \ref{fig:bbluminosity}). \citet{Farias2025} found that a sub-sample of SNe~Ibn exhibit a wide range of CSM masses ($\sim0.01-1$\,\Msun) and ejecta masses ($\sim0.1-10$\,\Msun), suggesting that many originate from a binary progenitor system. However, massive He-rich WR stars remain viable progenitors for SNe~Ibn with very high ejecta masses ($\sim10$\,\Msun), indicating there are multiple progenitor channels. Given that SN~2024abvb is photometrically similar to SNe~Ibn at early times, its pre-explosion progenitor and mass-loss mechanism may plausibly have been similar. The derived ejecta and CSM masses for SN~2024abvb (2.59\,\Msun and 0.3\,\Msun, respectively) are consistent with those of SNe~Ibn. However, the derived nickel mass is an order of magnitude higher than those reported in \citet{Farias2025}. Despite this, a binary system still remains a possible explanation for its progenitor system or at the very least a heavily stripped star exploding in a dense H/He-poor and C-rich CSM.

\subsubsection{Considerations for a SN~Icn}

The strong photometric diversity of SNe~Icn \citep{pellegrino_diverse_2022, Davis2023} indicates that SN~2024abvb falls within the observed range for this class. The six known SNe~Icn span a wide range of peak bolometric luminosities ($0.6-3\times10^{44}\text{erg s}^{-1}$) (top panel in Figure \ref{fig:bbluminosity}) and SN~2024abvb has a peak luminosity comparable to that of SN~2019hgp and a similar post-peak evolution to SN~2022ann. The early light-curve evolution (Figure \ref{fig:grmag_comparison}) among SNe~Icn is also diverse, with SN~2021csp showing a fast post-peak decline, while SN~2022ann fades more gradually. The light curve behaviour of SN~2024abvb falls well within this observed range. Additionally, SN~2024abvb maintains an approximately constant $g-r \approx 0$ colour during the first 10 days after maximum light, consistent with most of the SNe~Icn sample that have well-sampled colour coverage, such as SN~2019hgp and SN~2021csp. 

Near maximum light, SN~2024abvb displays spectral features similar to those of SNe~Icn. Its optical spectra $+1.2$ days after peak lack H and \He\ emission and show narrow \CII\, P~Cygni lines, matching SN~2012csp and SN~2022ann. These features suggest that SNe~Icn have similar composition and C-rich and H/He-poor circumstellar environments. In addition, SN~2024abvb, like SN~2022ann, shows no broad features after peak light or once the narrow emission lines begin to fade. \citet{Davis2023} attribute this to either a very low explosion velocity ($\sim800$\kms) or obscured SN ejecta. The low-velocity scenario is favoured here, as we see evidence of ongoing CSI with a C-rich CSM in our spectra. 

Based on its optical spectra, SN~2024abvb cannot strictly be classified as a SNe~Ibn or SNe~Icn. Although it shows early-time similarities to flash-ionised SNe~Ibn, the presence of strong \CII\, features, and the absence of \HeI\, emission at late times suggest that it is best interpreted as a SN~Icn with SN~Ibn-like characteristics. Placing SN~2024abvb in the context of transitional SN~Ibn/Icn events, we favour interpreting it as a transition from SN~Ibn to SN~Icn. If so, SN~2024abvb would represent the first example of such a transitional event, noting that \citet{Pursiainen2023} and \citet{Gangopadhyay2025B} previously suggested SN~2023emq and SN~2023xgo, respectively, may represent the opposite transition.

\subsection{Local environment and implications for the progenitor}\label{localenv}
To characterise the local environment of SN~2024abvb, we follow the methodology of \citet{Sand2018, Sand2021} to constrain the local SFR. Because SN~2024abvb is well separated from its host galaxy, the properties of the local stellar environment may provide important insights into the final stages of the progenitor system prior to explosion. A careful examination of Figure \ref{fig:spectra_timeseries} shows no H emission features. As such, we constrain the SFR of the local environment by estimating the maximum \Ha\, emission using the flux-calibrated and redshift-corrected late-phase spectrum obtained at +21.9\,d relative to the $o-$band maximum (see Figure \ref{fig:spectra_timeseries}). Previous works have predicted that if a SN progenitor system contains a non-degenerate companion star, interaction between the SN ejecta and the companion would produce narrow H emission lines with FWHM $\approx1000$\,\kms at late times (\citealp{Botyanzski2018, Tucker2020}, and others). We note that this approach entails systematic uncertainties; a more detailed treatment would require assumptions about the companion star type and, more challenging, the explosion energy and binary separation, which are beyond the scope of this work.    

Here, we set a quantitative limit on the maximum \Ha\, emission by implanting a synthetic emission line into our data. First, we bin the flux-calibrated and redshift-corrected spectrum to the spectral resolution of the data, $\sigma\approx5$\AA. This resolution is calculated from using the red channel grating with the medium slicer (R\,$\sim1800$) and a central wavelength, $\lambda\approx7150$. Note that \citet{Sand2021, Sand2018} used a nebular phase spectrum for their analysis, but since SN~2024abvb faded well before it reached the nebular phase, we use the last spectrum in our dataset (i.e., KCWI spectrum at +21.9\,d). We then employ a second-order Savitsky-Golay filter with a width of 180\,\AA~ to smooth the spectrum on scales greater than the expected emission (FWHM $\approx1000$\,\kms) in order to determine the continuum around the \Ha\, line wavelengths. This width was chosen as we found that smaller widths prevented us from detecting and measuring the implanted faint \Ha\, emission feature. The synthetic \Ha\, emission is measured from the residual spectrum, which is the difference between the smoothed and original spectrum. We assume a line width of FWHM $\approx1000$\,\kms~ and a peak flux that is four times the root mean square of the residual spectrum to be the maximum parameters for the implanted \Ha\, Gaussian feature to remain undetected. We obtain a 1$\sigma$ upper limit of the \Ha\, flux of $1.58\times10^{-16} \text{erg}~\text{s}^{-1}~\text{cm}^{-2}$, which has the luminosity limit of $5.63\times10^{38} \text{erg}~\text{s}^{-1}$ at a distance of $172$\,Mpc.

We also use the derived \Ha\, flux to trace the ongoing star-formation history (\lt 16 Myr old) of the local environment \citep{Gogarten2009, Dasitdar2025}. We employ Equation 2 from \citet{Kennicutt1998}, which relates the \Ha\, luminosity to the SFR and the scaling factor (0.63) from \citet{Madau2014}, to convert from the Salpeter IMF to the Chabrier IMF. From this, we calculate a local SFR limit of $0.0028~$\Msun\,yr$^{-1}$. This value is significantly smaller than the global SFR, i.e. $0.10~$\Msun\,yr$^{-1}$ (see Section \ref{sec:host-galaxy}), which is not surprising given the extreme offset between SN~2024abvb and its host galaxy. We note that this local SFR is consistent with two SNe~Ibn found highly offset from their host-galaxy: SN~2023tsz \citep{Warwick2025} and PS1-12sk \citep{Sanders2013}. For SN~2023tsz, \citet{Warwick2025} do not rule out a single stripped star progenitor given the exceptionally low SFR of the environment, and \citet{Hosseinzadeh2017} propose that the progenitor of PS1-12sk could be a massive star ejected or tidally stripped from a nearby ultra-compact dwarf galaxy. If an ultra-compact dwarf galaxy is present near SN~2024abvb, it would be extremely faint as pre-explosion images (limiting magnitude of $\sim$20 mag in Swope $r$-band observations \citealp{Coulter2017}) show no signs of any galaxies at the SN location. 

Additionally, we calculate the local SFR density. The KCWI spectrum was extracted using an aperture with a radius of 0.294\arcsec, corresponding to a physical radius of 0.245 kpc. This aperture subtends an area of of 0.19 kpc$^{2}$ and, when combined with the SFR upper limit derived above, yields a SFR density of $1.5\times10^{-2}\,\text{M}_{\odot}\,\text{yr}^{-1}\,\text{kpc}^{-2}$. Comparing our inferred SFR density with those of SNe~Ibn and SNe~IIn at the explosion sites presented in Figure 3 of \citet{Hosseinzadeh2019}, which were also derived from \Ha\, luminosities, we find that SN~2024abvb has a higher SFR density than 10 SNe~Ibn (59\% of the sample) and 9 SNe~IIn (57\% of the sample). This places SN~2024abvb above the median SFR density of both SNe~Ibn and SNe~IIn environments, suggesting that the local environment is broadly consistent with, though somewhat more star-forming than, the majority of the comparison sample.

We use the far ultraviolet (FUV) emission to trace the recent (16 -- 100 Myr) local star formation history. The FUV data yielded an upper limit of 20.8~mag. This value is an upper limit from our \emph{Swift} UV observation in the \textit{UVW2} band after using the package \emph{uvotimsum} and following the aperture photometry procedure outlined in Section \ref{follow-up-phot}. The FUV magnitude was corrected for the line-of-sight MW extinction. Using Equation 3 from \citet{Karachentsev2013}, we obtained log$_{10}$(SFR[M$_{\odot}$yr$^{-1}$]) = $-1.40\pm0.001$, equivalent to a SFR of $0.039\pm0.0001$ \Msun yr$^{-1}$. We note that this method yields a SFR value greater than the ongoing local SFR, which is likely due to the \textit{UVW2} band overlapping with the NUV band wavelength range leading to an overestimation of the recent local SFR. Nonetheless, this value is still smaller than the global SFR estimate of $0.10~$\Msun yr$^{-1}$ and provides an upper bound to the recent SFR in the local environment. Compared to the SFR distribution of SNe~Ibn and SESNe hosts (see Figure 10 in \citealt{Warwick2025}), the local SFR of SN~2024abvb lies in the bottom $\sim5$th percentile of SESNe hosts, suggesting a WR progenitor is unlikely. 

As proposed by \citet{pellegrino_diverse_2022}, a progenitor less massive than a WR star and residing in a binary system would have a longer evolutionary lifetime, thereby reducing the runaway velocity required to reach the observed location of SN~2024abvb. In this scenario, binary stripping would produce a substantial amount of CSM around the progenitor, rather than the line-driven mass loss that occurs during the main-sequence stage of a massive star. Furthermore, the low host-galaxy metallicity ($\log_{10}(\text{Z}_{*}/\text{Z}_{\odot}) = -1.38^{+0.38}_{-0.34}$) implies a low progenitor metallicity, making significant line-driven mass loss unlikely for SN~2024abvb. However, this metallicity estimate is inferred from photometric measurements alone and should be treated with caution. Another possibility is extreme episodic mass loss. For example, core-collapse progenitor stars are particular active (e.g., \citealp[]{Zenati2025, Killestein2025}), especially SNe~Ibn. The progenitor stars of SN~2006jc and SN~2019uo \citep{Foley2007, Pastorello2007, Strotjohann2021} showed luminous precursor outbursts $1-2$ years before explosion. No precursor activity brighter than $m_r \approx 20$ mag was reported for SN~2024abvb before its discovery, suggesting that the progenitor did not experience luminous eruptive episodes before collapse. Considering the observables of SN~2024abvb, including the ejecta mass ($\text{M}_{\text{ej}}$) and $^{56}$Ni mass ($\text{M}_{\text{Ni}}$) inferred from bolometric modelling, and the large projected offset of the SN from its host galaxy, we suggest the following plausible progenitor explanation: a binary system in which the primary star is of relatively low mass and undergoes significant mass loss through binary stripping before exploding within a dense, H/He-poor CSM.

\section{Conclusion} \label{sec:conclusion}
We have presented the UV, optical, and NIR photometric and spectroscopic observations of SN~2024abvb, a unique Type Icn SN exhibiting \CII\, emission in its early spectra and the largest projected offset from its host-galaxy reported to date (22.4\,kpc). SN~2024abvb is so far the sixth SNe~Icn discovered, providing a rare opportunity to investigate its progenitor avenues and explosion environment of this class. The key findings from our analysis are summarised here:

\begin{itemize}
    \item SN~2024abvb reaches a peak magnitude of $\text{M}_{r}=-19.55\pm0.11$, placing it among the most luminous members of the SNe~Icn class. The SN declines at a rate of 0.07 mag day$^{-1}$ in the $r$ band, which is consistent with other known SNe~Icn \citep{Fraser2021, Perley2022,pellegrino_diverse_2022, Davis2023} but slow compared to SNe~Ibn \citep{Hosseinzadeh2017}. 
    \item During the first week, SN~2024abvb shows evidence of interaction with a H/He-poor, C-rich CSM in its optical spectra. Its early absence of \CIII\, $\lambda5700$ could suggest a Type~Ibn classification, but the persistent lack of \He\, emission at late times indicates a transition from SN~Ibn to SN~Icn with low C ionisation, making SN~2024abvb the first known example of such an event. 
    \item SN~2024abvb was found 22.4\,kpc away from its host galaxy, which is the largest distance of the SNe~Icn sample and the third largest of the SNe~Ibn sources measured so far. This offset is similar to SN~Ibn PS1-12sk, which also exploded in a region devoid of star formation. These atypical explosion environments suggest the possibility of a progenitor channel other than a direct collapse of WR stars, or WR-like massive stars, which are typically favoured for SNe~Ibn/Icn. 
    \item The host-galaxy of SN~2024abvb is an intermediate-mass galaxy with a stellar mass of $\log_{10}(\mathrm{M}_{*}/\mathrm{M}_{\odot}) = 9.66^{+0.12}_{-0.12}$ and an extremely low sSFR of $\log_{10}(\text{sSFR}$yr$^{-1}) = -10.62^{+0.79}_{-1.22}$. The low metallicity ($\log_{10}(\text{Z}_{*}/\text{Z}_{\odot}) = -1.38^{+0.38}_{-0.34}$), could reflect a metal-poor progenitor for SN~2024abvb, potentially limiting the role of line-driven winds in stripping its H/He envelope. This would suggest a scenario in which a low-mass star was stripped by a compact companion and exploded within a dense, H/He-poor CSM. 
    \item We fit the combined UV, optical, and NIR light curves of SN~2024abvb with a radioactive decay $+$ CSM interaction model, and find values of $\text{M}_{\text{ej}}\sim$2.59$\,\text{M}_{\odot}$, $\text{M}_{\text{CSM}}\sim$0.28$\,\text{M}_{\odot}$, and $\text{M}_{\text{Ni}}\sim$0.1$\,\text{M}_{\odot}$, all of which are consistent with rapidly evolving SNe~Ibn \citep{Pellegrino2022}. However, other SNe~Icn show lower $\text{M}_{\text{Ni}}$. This diversity may arise from differences in how physical parameters are inferred or assumed for different events. 
    \item From implanting a synthetic \Ha\, feature, we derive a local SFR density of $1.5\times10^{-2}\,\text{M}_{\odot}\,\text{yr}^{-1}\,\text{kpc}^{-2}$, consistent with typical SNe~Ibn and SNe~IIn hosts \citep{Hosseinzadeh2019}. \textit{Swift} UV yield a local SFR of 0.039 $\pm$ 0.0001, placing it in the bottom $\sim5$th percentile of SESNe host galaxies. 
\end{itemize}


\section{Acknowledgements}
W.B.H. acknowledges support from the National Science Foundation Graduate Research Fellowship Program under Grant No. 2236415. 
M.D.S. is funded by the Independent Research Fund Denmark (IRFD, grant number  10.46540/2032-00022B) 
L.G. acknowledges financial support from CSIC, MCIN and AEI 10.13039/501100011033 under projects PID2023-151307NB-I00, PIE 20215AT016, and CEX2020-001058-M.
J.T.H acknowledges support provided by NASA through the NASA Hubble Fellowship grant HST-HF2-51577.001-A awarded by the Space Telescope Science Institute, which is operated by the Association of Universities for Research in Astronomy, Incorporated, under NASA contract NAS5-26555.  D.O.J. acknowledges support from NSF grants AST-2407632, AST-2429450, and AST-2510993, NASA grant 80NSSC24M0023, and HST/JWST grants HST-GO-17128.028 and JWST-GO-05324.031, awarded by the Space Telescope Science Institute (STScI), which is operated by the Association of Universities for Research in Astronomy, Inc., for NASA, under contract NAS5-26555.

Based in part on data acquired at the ANU 2.3-metre telescope, under proposal ID: 2425062, 2425126, 2425128. The automation of the telescope was made possible through an initial grant provided by the Centre of Gravitational Astrophysics and the Research School of Astronomy and Astrophysics at the Australian National University and through a grant provided by the Australian Research Council through LE230100063. The Lens proposal system is maintained by the AAO Research Data \& Software team as part of the Data Central Science Platform. We acknowledge the traditional custodians of the land on which the telescope stands, the Gamilaraay people, and pay our respects to elders past and present.

The authors wish to recognise and acknowledge the significant cultural role and reverence that the summit of Maunakea has always had within the indigenous Hawaiian community. We are most fortunate to have the opportunity to conduct observations from this mountain.

Some of the data presented herein were obtained at the Keck Observatory, which is a private 501(c) (3) nonprofit organization operated as a scientific partnership among the California Institute of Technology, the University of California, and the National Aeronautics and Space Administration. The Observatory was made possible by the generous financial support of the W.\ M.\ Keck Foundation.

UKIRT is owned by the University of Hawaii (UH) and operated by the UH Institute for Astronomy.

The Infrared Telescope Facility, which is operated by the University of Hawaii under contract 80HQTR24DA010 with the National Aeronautics and Space Administration.

Parts of this research were supported by the Australian Research Council Centre of Excellence for Gravitational Wave Discovery (OzGrav), through project number CE230100016.

B.E.T. acknowledges support from the Kavli Foundation.

IMACS data were obtained while A.P. was supported by a Carnegie Fellowship through the Observatories of the Carnegie Institute for Science.

\bibliography{ref}

\renewcommand{\thefigure}{A\arabic{figure}}
\renewcommand{\thetable}{A\arabic{table}}

\setcounter{figure}{0}
\setcounter{table}{0}
\begin{appendix}
\section{Host Galaxy}
\begin{figure}
    \centering
    \includegraphics[width=\linewidth]{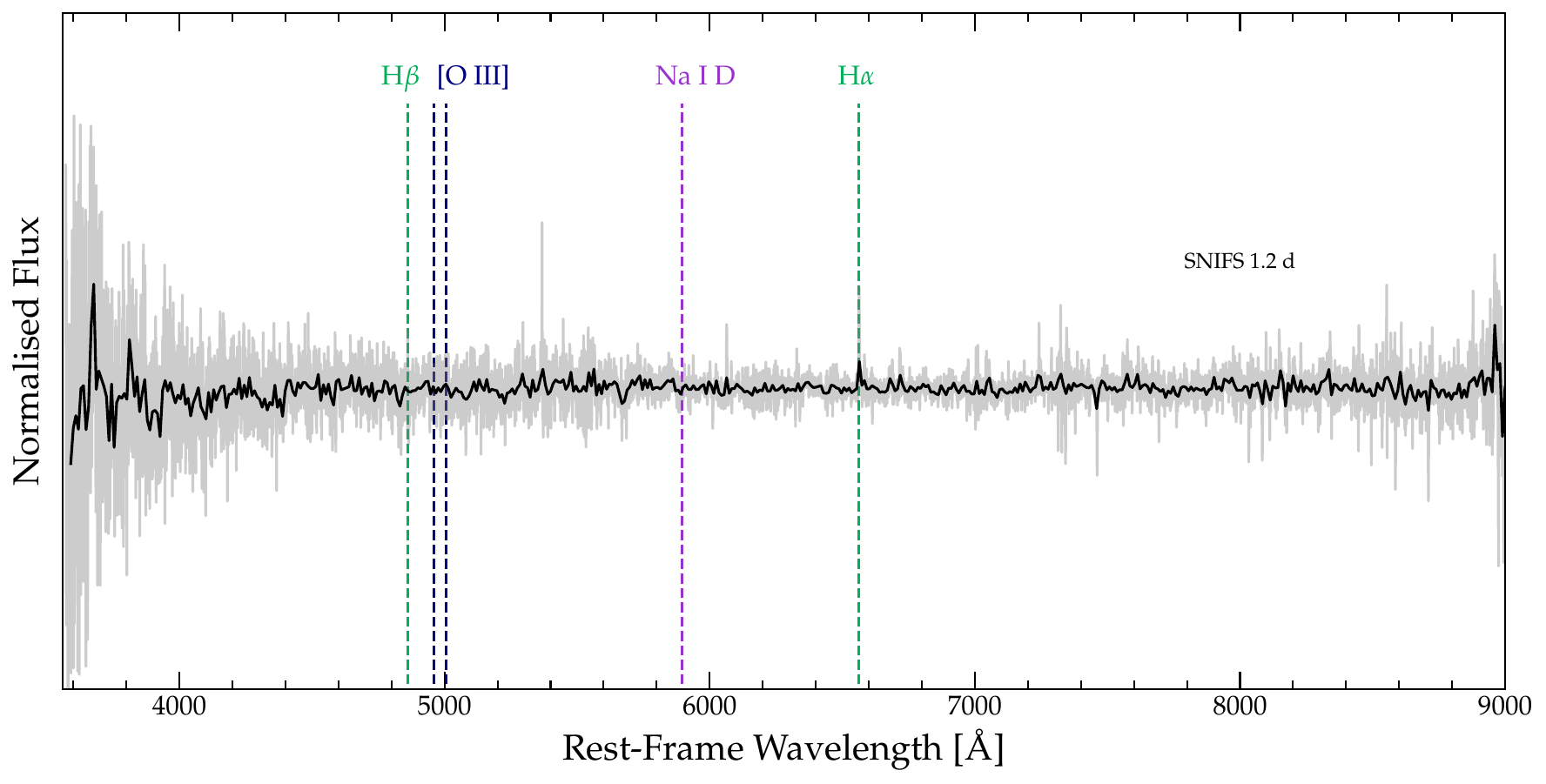}
    \vspace{-2mm}
    \caption{Host galaxy spectrum of SN~2024abvb taken 1.2 days after maximum $o$-band brightness. We find weak 
    \Ha, \Hb, and [\OIII] emission lines. The lack of \ion{Na}{i}~D absorption indicates that there is minor or no host-galaxy extinction present.} \label{fig:hostgal}
    \vspace{-2mm}
\end{figure}    

\section{Log of spectroscopic observations}
In Table \ref{tab:spectra_table}, we present the optical and NIR spectroscopic observations of SN~2024abvb.
\begin{table}[]
    \centering
    \caption{Log of spectroscopic observations for SN~2024abvb.}\label{tab:spectra_table}
    \begin{tabular}{ccccc}
        \hline 
        UT Date & MJD &  Telescope & Instrument & Range [\AA]\\ 
        \hline \hline 
        \multicolumn{5}{c}{\emph{Optical}} \\
        \hline
            2024-11-23 & 60637.4 & UH 2.2m          & SNIFS & 3280-8750\\ 
            2024-11-26 & 60640.4 & UH 2.2m          & SNIFS & 3280-8750\\ 
            2024-11-27 & 60641.9 & NOT           & ALFOSC & 3460-9300\\ 
            2024-12-02 & 60646.4 & ANU 2.3m         & WiFeS & 3920-8680\\ 
            2024-12-02 & 60646.9 & NOT            & ALFOSC & 3530-9330\\ 
            2024-12-05 & 60649.4 & ANU 2.3m         & WiFeS  & 3920-8590\\ 
            2024-12-08 & 60652.4 & ANU 2.3m         & WiFeS & 3920-8590\\ 
            2024-12-10 & 60654.1 & Magellan-Baade   & IMACS & 4090-9010\\ 
            2024-12-16 & 60660.5 & ANU 2.3m         & WiFeS & 3920-8590\\ 
            2024-12-19 & 60663.2 & UH 2.2m          & SNIFS & 3280-8750\\ 
            2024-12-22 & 60666.3 & UH 2.2m          & SNIFS & 3280-8750\\
            2024-12-23 & 60667.2 & Keck II          & KCWI & 3160-8560\\ 
\vspace{0.50mm} \\
        \hline 
        \multicolumn{5}{c}{\emph{NIR}} \\
        \hline
            2024-12-08 & 60652.0 & Magellan-Baade & FIRE & 7500-23300\\ 
            2024-12-18 & 60662.3 & SpeX & IRTF & 6800-24300\\ 

 \vspace{0.50mm} \\
            \hline \\
    \end{tabular}
\end{table}

\section{Radioactive Decay Model} \label{radioactivedecaymodel}

The radioactive decay model assumes that the light curve is powered by the decay chain $^{56}$Ni to $^{56}$Co to $^{56}$Fe. We fit the analytical model of \citet{Arnett1982} to the UVOIR multiband light curve, assuming homologous expansion with a uniform density profile for the ejecta.
The optical (electron scattering) opacity was fixed to $\kappa_o = 0.04~\text{cm}^{2}~\text{g}^{-1}$ following \citet{pellegrino_diverse_2022}, and the gamma opacity was fixed to $\kappa_\gamma = 0.08~\text{cm}^{2}~\text{g}^{-1}$ \citep{Rabinak2011}. We impose the physical constraint that $\text{f}_{\text{Ni}}$ does not exceed the total ejecta mass. The free parameters in the fit include the nickel fraction $f_{\text{Ni}}$, the ejecta mass M$_{\text{ej}}$, ejecta velocity V$_{\text{ej}}$, and the temperature floor T$_{\text{floor}}$.

\begin{figure*}[hbt!]
  \centering
  \begin{minipage}{0.45\textwidth}
    \centering
    \includegraphics[width=\linewidth]{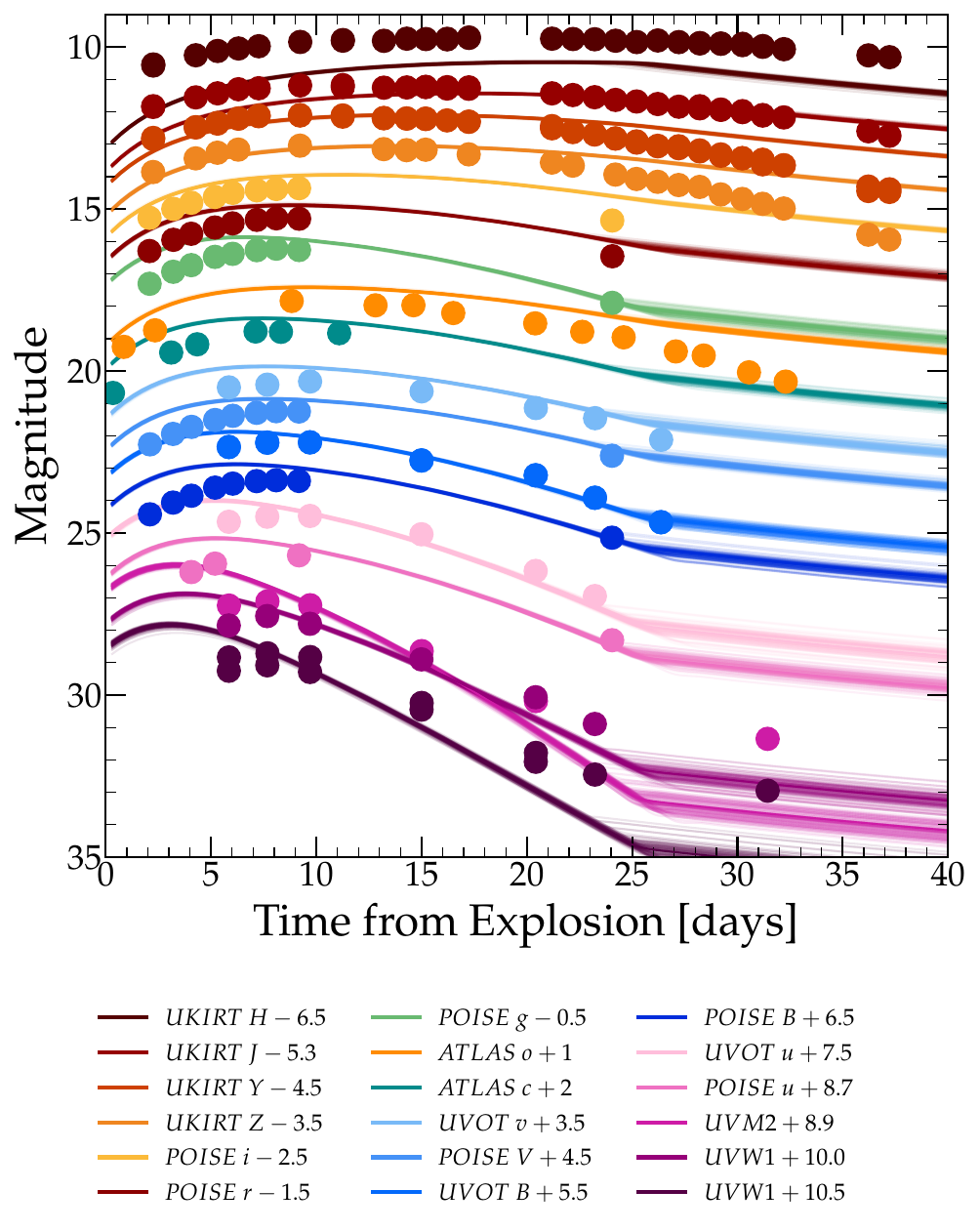}
  \end{minipage}
    \hspace{0.001\textwidth}
  \begin{minipage}{0.45\textwidth}
    \centering
    \includegraphics[width=\linewidth]{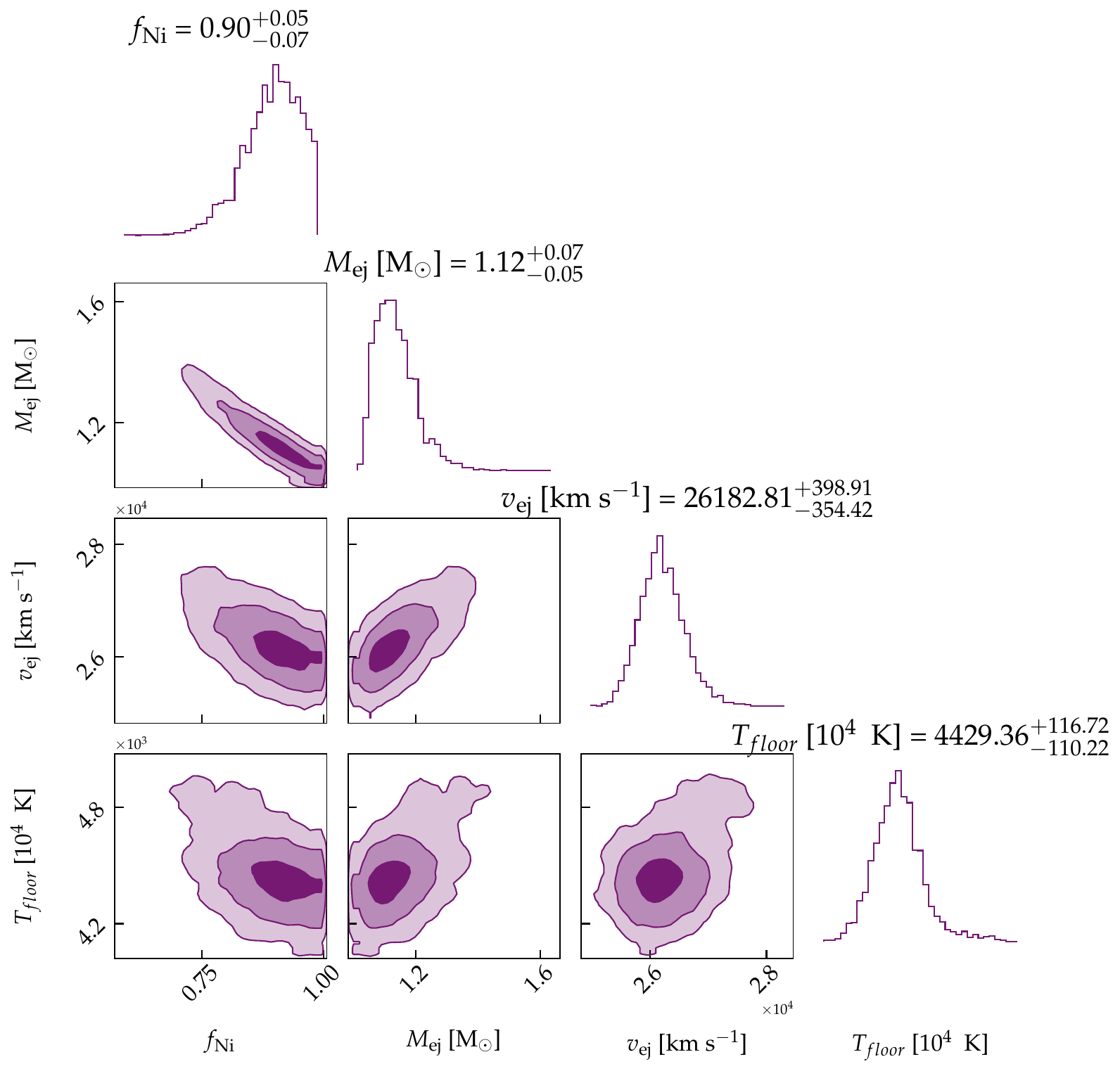}
  \end{minipage}
  \caption{\textit{Left:} The best-fit multi-band light curves of SN~2024abvb using the Arnett model \citep{Arnett1982}. \textit{Right:} The corner plot of the Arnett model fit to the multi-band light curve of SN~2024abvb in \texttt{Redback}.}
  \label{fig:redbackarnett}
\end{figure*}

The best-fit light curve and corresponding corner plot are shown in Figure \ref{fig:redbackarnett} and the derived parameters are presented in Table \ref{tab:arnett_table}. Overall, the radioactive decay model is a reasonable fit to the multiband light curves; however, it overestimates the rise to peak ($0-10$ days from explosion), particularly in the optical and UV bands, and slightly over-fits the decline rate. A $^{56}$Ni mass of  $\text{M}_{\text{Ni}}~\text{of}~\simeq 1.01~\text{M}_{\odot}$ ($>$~90\%) is required to reproduce a reasonable fit to the light curve, which is unusually high for of CCSNe. By comparison, \citet[]{Anderson2019} report a median  $\text{M}_{\text{Ni}}<$0.4\,$\text{M}_{\odot}$ across all CCSNe types, the derived value also exceeds the high-end distribution of SESNe \citep{Lyman2016}. Relative to other SNe~Icn, SN~2024abvb is an order of magnitude greater (see Figure 8 in \citet{pellegrino_diverse_2022}). A lower limit on the $^{56}$Ni mass can be estimated from the late-time bolometric luminosity using the relation of \citet{Arnett1982}, as applied by \citet{Davis2023}. Assuming instantaneous $\gamma$-ray trapping and that SN~2024abvb is powered by $^{56}$Ni at late times, the measured bolometric luminosity ($\sim2\times10^{41}\,\text{erg s}^{-1}$) implies a $^{56}$Ni mass of $\approx0.007$ \Msun, which is smaller than typical values for SESNe \citep{Lyman2016, Afsariardchi2021}. From the bolometric light curve, we also estimate an ejecta velocity ($v_{ej}$) of $\sim26000$\kms, higher but comparable in order of magnitude to the photospheric velocity ($\sim18000$\kms). This is similar to SN~2021csp, for which \citet{pellegrino_diverse_2022} derived $v_{ej} \sim23000$\kms. However, the lack of spectral features exceeding $>2000$\kms presents a clear discrepancy. As such, these results indicate that radioactive decay alone is unlikely to be the dominant power source of SN~2024abvb. 

\begin{table}[]
    \centering
    \caption{Best-fit parameters for the Arnett model as described in Section \ref{radioactivedecaymodel}.}\label{tab:arnett_table}
    \begin{tabular}{lll}
        \hline
        Parameters & Priors &  Best-Fitted Values  \\
        \hline \hline
        \\    
            $\text{f}_{\text{Ni}}$ &  $\mathcal{U}~[0.01, 0.99]$    & $0.90_{-0.07}^{+0.05}$      \vspace{0.50mm} \\
            $\text{M}_{\text{ej}}~[\text{M}_{\odot}]$ & log$~\mathcal{U}~[0.01, 2] $  & $1.12_{-0.05}^{+0.07}$         \vspace{0.50mm} \\
            $\text{V}_{\text{ej}}~[\text{km~s}^{-1}]$ & $\mathcal{U}~[10^{2}, 5\times10^{4}]$  & $26182.81_{-354.42}^{+398.91}$      \vspace{0.50mm} \\
            $\text{T}_{\text{floor}}~[\text{K}]$ &  $\mathcal{U}~[10^{3}, 10^{4}]$ & $4429.36_{-110.22}^{+116.72}$ \\       
 \vspace{0.20mm} \\
            \hline \\
    \end{tabular}
\end{table}

\end{appendix}

\end{document}